\documentstyle[12pt]{article}
{\baselineskip 12pt}
\begin{document}
\begin{flushright}
{\bf WSP--IF 99--55 \\
May 10, 1999}
\end{flushright}
\begin{center}
{\Large \bf A contribution to the discussion \\   of the 
matter--antimatter asymmetry problem} \\
\hfill \\ K. Urbanowski$^{\ast}$ \\  \hfill  \\
Pedagogical University, Institute of Physics, \\
Plac Slowianski 6, 65-069 Zielona Gora, Poland. \\
\hfill \\
May 10, 1999 
\end{center}
{\noindent}{\em PACS numbers:} 98.80.Cq., 98.80.Bp, 11.30.Er., 
13.20.Eb. \\
{\em Keywords:} Matter--antimatter asymmetry; Particle--antiparticle 
masses; Unstable particles; CPT invariance \\
\nopagebreak
\begin{abstract}
Using a more accurate effective Hamiltonian governig the time 
evolution in the particle--antiparticle subspace of states 
than the one obtained within the Lee--Oehme--Yang  approach we show that 
in the case of particles created at the instant $t = t_{0}$ only 
the masses of a stable particle and its antiparticle are the same at 
all $t \geq t_{0}$ in a CPT invariant system, whereas the masses of an 
unstable particle and its antiparticle are equal only at $t = t_{0}$ 
and then during their time evolution they become 
slightly different for times $t \gg t_{0}$ if CP symmetry is violated 
but CPT symmetry holds. This property is used to show that if 
the baryon number B is not conserved then the asymmetry between numbers 
of unstable baryons and antibaryons can arise in a CPT invariant system
at $t \gg t_{0}$ even in the thermal equilibrium state of this 
system.
\end{abstract}  
$^{\ast}$e--mail: kurban@magda.iz.wsp.zgora.pl ,
kurban@omega.im.wsp.zgora.pl
\pagebreak[4]

\def\Tr{{\rm Tr \,}}

\section{Introduction.}

The knowledge of all the subtleties of the difference between a particle 
and its antiparticle has a fundamental meaning for understanding 
our Universe, and especially for the explanation of the problem why 
the observed Universe contains (acording to the physical and 
astrophysical data) an excess of matter over antimatter.
Searching for properties of the unstable particle--antiparticle pairs 
one usually uses an effective nonhermitean Hamiltonian \cite{1} ---  
\cite{14}, say $H_{\parallel}$, which in general can depend on time 
$t$ \cite{horwitz}, 
\begin{equation}
H_{\parallel} \equiv M - \frac{i}{2} \Gamma, \label{new1}
\end{equation}
where
\begin{equation}
M = M^{+}, \; \; \Gamma = {\Gamma}^{+}, \label{new1a}
\end{equation}
are $(2 \times 2)$ matrices, acting in a two--dimensional subspace 
${\cal H}_{\parallel}$ of the total state space $\cal H$ and $M$ is 
called the mass matrix, $\Gamma$ is the decay matrix \cite{1} --- 
\cite{5}. The standard method of derivation of such $H_{\parallel}$ 
bases on a modification of the Weisskopf--Wigner (WW) approximation 
\cite{ww}. Lee, Oehme and Yang (LOY) have adapted WW aproach to the 
case of a two particle subsystem \cite{1} --- \cite{4} to obtain their 
effective Hamiltonian $H_{\parallel} \equiv H_{LOY}$. Almost all 
properties of the neutral kaon complex, or another 
particle--antiparticle 
subsystem can be described by solving the Schr\"{o}dinger--like 
evolution equation \cite{1} --- \cite{14}, \cite{9} --- \cite{tsai} 
(we use $\hbar = c = k = 1$ units)
\begin{equation}
i \frac{\partial}{\partial t} |\psi ; t >_{\parallel} =
H_{\parallel} |\psi ; t >_{\parallel}
\label{l1}
\end{equation}
with the initial condition \cite{tsai,10}
\begin{equation}
{\parallel} \, |{\psi} ; t = t_{0} >_{\parallel} {\parallel} = 1, 
\; \; | {\psi}; t < t_{0} >_{\parallel} = 0, \label{init}
\end{equation}
for $| \psi ; t >_{\parallel}$ belonging to the subspace
${\cal H}_{\parallel} \subset {\cal H}$ spanned, e.g., by 
orthonormal neutral  kaons states $|K_{0}>, \; |{\overline{K}}_{0}>$, 
and so on, (then states corresponding with the decay products belong 
to ${\cal H} \ominus{\cal H}_{\parallel} \stackrel{\rm def}{=} 
{\cal H}_{\perp}$).
 
Solutions of Eq. (\ref{l1}) can be written in matrix  form  and
such  a  matrix  defines  the evolution    operator
(which    is     usually     nonunitary) $U_{\parallel}(t)$ acting in 
${\cal H}_{\parallel}$:
\begin{equation}
|\psi ; t >_{\parallel} = U_{\parallel}(t)
|\psi ;t_{0} = 0 >_{\parallel} \stackrel{\rm def}{=}
U_{\parallel}(t) |\psi >_{\parallel}, \label{l1a}
\end{equation}
where,
\begin{equation}
|\psi >_{\parallel} \equiv a_{1}|{\bf 1}> + a_{2}|{\bf 2}>, 
\label{l1b}
\end{equation}
and $|{\bf 1}>$ stands for vectors of the   $|K_{0}>,  \; 
|B_{0}>$, etc., type and $|{\bf 2}>$ denotes antiparticles  of  the  
particle "1": $|{\overline{K}}_{0}>, \; {\overline{B}}_{0}>$, and so 
on, $<{\bf j}|{\bf k}> = {\delta}_{jk}$, $j,k =1,2$. In many papers it 
is assumed that the real parts, $\Re (.)$, of diagonal matrix elements 
of $H_{\parallel}$:
\begin{equation}
\Re \, (h_{jj} )
\equiv M_{jj}, \; \;(j =1,2),
\label{m-jj}
\end{equation}
where
\begin{equation}
h_{jk}  =  <{\bf j}|H_{\parallel}|{\bf k}>, \; (j,k=1,2),
\label{h-jk}
\end{equation}
correspond to the masses  of particle "1" and its antiparticle "2" 
respectivelly \cite{1} --- \cite{leonid} (and such an interpretation 
of $\Re \, (h_{11})$ and $\Re \, (h_{22})$ will be used in this 
paper), whereas imaginary parts, $\Im (.)$,
\begin{equation}
-2 \Im \, (h_{jj}) \equiv {\Gamma}_{jj}, \; \;(j =1,2),
\label{g-jj}
\end{equation}
are interpreted as the decay widths of these particles \cite{1} --- 
\cite{leonid}. 

Physicists look for the mechanism generating the observed 
baryon--anti\-ba\-ryon asymmetry. Such a mechanism is necessary for an 
explanation of the matter--antimatter asymmetry in the Universe. 
Most theories of dynamic generation of the baryon assymetry
use the so--called Sakharov conditions \cite{m-antim} ---
\cite{m-antim11}. 
These conditions do not require CPT symmetry to be violated.
The existence of C and CP violating processes, the violation of the 
baryon number and the existence of nonequilibrium processes are 
sufficient for baryogenesis \cite{m-antim} --- \cite{m-antim11}. These 
conditions can be met in many papers describing different ways of 
generating the baryon asymmetry. Unfortunately, there is no answer to 
the question which one of them, if any, is corret. Maybe for this 
reason some models of processes producing matter--antimatter masses 
difference, in which a small violation of CPT symmetry at the origin is 
assumed, are also considered \cite{m-antim11} --- \cite{m-antim1a}. 
Such models are discussed in spite of the fact that  CPT symmetry is 
a fundamental theorem, called the CPT Theorem, of axiomatic quantum 
field theory which follows from locality, Lorentz invariance, and 
unitarity \cite{cpt}. This illustrates the importance of the problem 
of finding the mechanism producing the matter--antimatter mass 
asymmetry.

One consequence of the CPT theorem is that under the product of 
operations ${\cal C}, \, {\cal P}$ and $\cal T$ the total Hamiltonian 
$H$ of the system considered must be invariant. From this property and 
from  the properties of the LOY theory of time evolution in the subspace 
of states of two particle subsystem  prepared at some initial instant  
$t_{0}$ and then evolving in time $t > t_{0}$ \cite{1}, one usually 
infers that particles and antiparticles have exactly the same mass. 
Properties of singularities of scattering amplitudes appearing in 
S--matrix theory provide one with reasons for a similar conclusion. 
Generally, such a conlusion is considered to be the obvious. 

The aim of this paper is to show by using a more accurate approximation 
than the LOY theory, that only the masses of a stable particle and its 
antiparticle are equal in the CPT--invariant system whereas the masses 
of an unstable particle and its simultaneously created at $t = t_{0}$  
antiparticle must be slightly different at $t \gg t_{0}$ when CPT 
symmetry holds but CP symmetry does not. This effect is found by 
analysing only the properties of approximate solutions of 
the Schr\"{o}dinger equation for the initial conditions of type 
(\ref{init}), (\ref{l1b}). The paper is organized as follows. Formulae 
appearing in the LOY approach and the derivation of the more accurate 
effective Hamiltonian $H_{\parallel}$ than $H_{LOY}$ are described in 
short in Sec. 2. The properties of matrix elements of the 
$H_{\parallel}$ 
implied by CPT symmetry are discussed in Sec. 3. Possibilities of an 
experimental verification of the results obtained in Sec. 3 are 
considered in Sec. 4.  Sec. 5 contains a general discussion of 
the relations obtained in Sec. 3.,  and also some attempts
to show that 
the mechanism following from the final relation of  Sec. 3 
can generate
a conribution to the observed matter--antimatter asymmetry.
A summary and final 
remarks can be found in Sec. 6.

\section{Preliminaries.}

\subsection{$H_{LOY}$ and CPT--symmetry.}
Now, let us consider briefly some properties of the LOY model. 
Let $H$ be the total (selfadjoint) Hamiltonian, acting in $\cal H$ --- 
then  the  total unitary evolution  operator$U(t)$  fulfills  the  
Schr\"{o}dinger equation
\begin{equation}
i \frac{\partial}{\partial t} U(t)|\phi > =
H U(t)|\phi >,  \; \; U(0) = I,
\label{l2}
\end{equation}
where $I$ is the unit operator in $\cal H$, $|\phi > \equiv
|\phi ; t_{0} = 0> \in {\cal H}$  is  the  initial  state  of  the
system:
\begin{equation}
|\phi  >  \equiv  |\psi  >_{\parallel},  \label{l2a}
\end{equation}
$t \geq t_{0} = 0$, and, in  our case  $|\phi ;t> = U(t) |\phi >$. 
Let $P$ denote the projection operator onto the subspace 
${\cal H}_{\parallel}$: 
\begin{equation}
P{\cal H} =
{\cal H}_{\parallel}, \; \; \; P = P^{2} = P^{+}, 
\label{new2}
\end{equation}
then the subspace of decay products ${\cal H}_{\perp}$ equals
\begin{equation}
{\cal H}_{\perp}  = (I - P) {\cal H} 
\stackrel{\rm def}{=} Q {\cal H}, \; \; \; Q \equiv I - P.
\label{l7c}
\end{equation}
For the  case of neutral  kaons  or  neutral  $B$--mesons,  etc.,  
the projector $P$ can be chosen as follows:
\begin{equation}
P \equiv |{\bf 1}><{\bf 1}| + |{\bf 2}><{\bf 2}|.
\label{l3}
\end{equation}
We assume that the time independent basis vectors $|K_{0}>$ and 
$|{\overline{K}}_{0}>$ are defined analogously to the corresponding 
vectors used in the LOY theory of time evolution in neutral kaon complex 
\cite{1}: Vectors  $|K_{0}>$ and $|{\overline{K}}_{0}>$  can be 
identified with the eigenvectors of the so--called free Hamiltonian
$H^{(0)}  = H - H_{I}$, where $H_{I}$ denotes the interactions which are 
responsible for transitions between eigenvectors of $H^{(0)}$, i.e., 
for the decay process. 

In the LOY approach it is assumed that vectors $|{\bf 1}>$,  
$|{\bf 2}>$ considered above are the eigenstates of $H^{(0)}$ for a 
2-fold degenerate eigenvalue $m_{0}$:
\begin{equation}
H^{(0)} |{\bf j} > = m_{0} |{\bf j }>, \; \;  j = 1,2 . \label{b1}
\end{equation}
This means that
\begin{equation}
[P, H^{(0)}] = 0. \label{new3}
\end{equation}

The condition guaranteeing the occurence of transitions  between  
subspaces ${\cal H}_{\parallel}$ and ${\cal H}_{\perp}$, i.e.,  a  
decay process of states in ${\cal H}_{\parallel}$, can  be  written
as follows
\begin{equation}
[P,H_{I}] \neq 0 . \label{r32}
\end{equation}

Usually, in LOY and related approaches, it is assumed that
\begin{equation}
{\Theta}H^{(0)}{\Theta}^{-1} = {H^{(0)}}^{+} \equiv H^{(0)} , 
\label{r31}
\end{equation}
where $\Theta$ is the antiunitary operator:
\begin{equation}
\Theta \stackrel{\rm def}{=} {\cal C}{\cal P}{\cal T}.
\label{new4}
\end{equation}

Relation (\ref{r31}) is a particular form of the general 
transformation rule \cite{cpt} --- \cite{15}:
${\Theta} {\cal O} {\Theta}^{-1} \stackrel{\rm def}{=} 
{\cal O}_{CPT}^{+}$, where $\cal O$ is an arbitrary linear operator. 
Basic properties of anti--linear and linear operators, their
products and commutators are described, eg., in 
\cite{6,cpt,messiah,bohm}. 
 
The subspace of neutral kaons ${\cal H}_{\parallel}$ is assumed to
be invariant under $\Theta$:
\begin{equation}
{\Theta} P {\Theta}^{-1} =  P. \label{9aa}
\end{equation}

In the kaon rest frame, the time evolution for $t \geq t_{0} =0$ is 
governed by the Schr\"{o}dinger equation (\ref{l2}) with the initial 
condition (\ref{init}),  where the initial state of the system 
has the form (\ref{l2a}), (\ref{l1b}). Within assumptions 
(\ref{b1}) --- (\ref{r32}) and assuming the following form of 
$| \psi ; t >_{\parallel}$ for $t \geq t_{0}$ (see \cite{1}, formula 
(21)),
\begin{equation}
| \psi ; t>_{\parallel} = e^{- \frac{\lambda}{2}t} 
|\psi >_{\parallel} , 
\label{form}
\end{equation}
(where $\lambda$ is a complex number, $ \Re \, ({\lambda}) > 0$),
the  Weisskopf--Wigner approach leads to the 
following formulae for the matrix elements $h_{jk}^{LOY}$ of $H_{LOY}$  
(see, e.g., \cite{1,2,3,5,improved}):
\begin{eqnarray}
h_{jk}^{LOY} & = &  <{\bf j}|H_{LOY}|{\bf k}> 
=  H_{jk} - {\Sigma}_{jk} (m_{0} )  \label{b5} \\
& = & M_{jk}^{LOY} - \frac{i}{2} {\Gamma}_{jk}^{LOY},  
\; \; \; (j,k = 1,2), \label{b5a}
\end{eqnarray}
where, in this case,
\begin{eqnarray}
H_{jk} &\equiv&  <{\bf j} |PHP| {\bf k} >  = <{\bf j} |H| {\bf k} > 
\nonumber \\
&\equiv& <{\bf j} |(H^{(0)} + H_{I} )| {\bf k} >
\equiv m_{0} {\delta}_{jk} + <{\bf j}|H_{I}|{\bf k}> ,
\label{b6}
\end{eqnarray}
${\Sigma}_{jk} ( \epsilon ) = < {\bf j} \mid \Sigma
( \epsilon ) \mid {\bf k} >$ and
\begin{equation}
\Sigma ( \epsilon ) = PHQ \frac{1}{QHQ - \epsilon - i 0} QHP
\stackrel{\rm def}{=} {\Sigma}^{R}(\epsilon ) + 
i {\Sigma}^{I}(\epsilon ),   \label{r24}
\end{equation}
\begin{eqnarray}
{\Sigma}^{R}(\epsilon ) & = & PHQ \, {\bf P} \frac{1}{QHQ - \epsilon}
QHP, \label{r24a} \\
{\Sigma}^{I}(\epsilon ) & = & \pi PHQ \delta (QHQ - \epsilon ) QHP,
\label{r24b}
\end{eqnarray}  
and for real $\epsilon$, ${\Sigma}^{R}(\epsilon ) = 
{\Sigma}^{R}(\epsilon )^{+}$ and ${\Sigma}^{I}(\epsilon ) = 
{\Sigma}^{I}(\epsilon )^{+}$.  (In Eq. (\ref{r24a}) {\bf P} denotes 
principal value). 

Now, if ${\Theta}H_{I}{\Theta}^{-1} = H_{I}$,  
then using, e.g., the following phase convention \cite{2} --- \cite{6}
\begin{equation}
\Theta |{\bf 1}> \stackrel{\rm def}{=} - |{\bf 2}>, \;\; 
\Theta|{\bf 2}> 
\stackrel{\rm def}{=} - |{\bf 1}>, \label{cpt1}
\end{equation}
and taking into account that $< \psi | \varphi > = 
<{\Theta}{\varphi}|{\Theta}{\psi}>$,
one easily finds from (\ref{b5}) -- (\ref{r24b})  that
\begin{equation}
h_{11}^{LOY} = h_{22}^{LOY}  \label{b8}
\end{equation}
in the CPT--invariant system. This  is  the standard result of the 
LOY approach and this is the picture  which one meets in  the  
literature  \cite{1}  --- \cite {baldo},  \cite{chiu}. So, within 
this approximation the masses of particle "1" and its antiparticle 
"2" are equal in the system preserving CPT symmetry,
\begin{equation}
M_{11}^{LOY} = M_{22}^{LOY}. \label{m-loy}
\end{equation}

\subsection{Beyond the LOY approximation.}
The approximate formulae for $H_{\parallel}(t)$ (where $t \geq 0$ ) 
have been  derived  in \cite{9,10} assuming that
\begin{equation}
[P,H] \neq 0,
\label{l7}
\end{equation} 
and using  the  Kr\'{o}likowski--Rzewuski (KR) equation for the 
projection of a state vector \cite{7,pra}. Such an approach is 
convenient 
when one searches for the properties of the systems prepared at some 
initial instant $t_{0}$, say, $t_{0} = 0$, and then evolving in time 
$t \geq t_{0} = 0$ \cite{pra}. In such cases the S--matrix formalism 
mentioned at the end of Sec.1 does not work (it works when time $t$ 
varies from $t=- \infty$ to $t=+ \infty$), and all conclusions 
following from the scattering theory need not be true for these systems.

The KR Equation results from the Schr\"{o}dinger equation (\ref{l2}) 
for the  total system under consideration, and, in the case  of initial 
conditions of the type (\ref{l2a}), takes the following form 
\begin{equation}
( i \frac{\partial}{ {\partial} t} - PHP ) U_{\parallel}(t)
=  - i \int_{0}^{\infty} K(t - \tau ) U_{\parallel}
( \tau ) d \tau,  \label{B1}
\end{equation}
where $ U_{\parallel} (0)  =  P$ and $ U_{\parallel} (t < 0)  =  0$ 
\cite{10},
\begin{equation}
K(t)  =  {\mit \Theta} (t) PHQ \exp (-itQHQ)QHP,
\label{B2}
\end{equation}
and ${\mit \Theta} (t)  =  { \{ } 1 \;{\rm for} \; t 
\geq 0, \; \; 0 \;{\rm for} \; t < 0 { \} }$. On the other hand, in 
this case simply
\begin{equation}
U_{\parallel}(t) \equiv PU(t)P = Pe^{-itH}P,
\; \; (t \geq t_{0}). \label{u-par}
\end{equation}

Defining
\begin{equation}
H_{\parallel}(t) \stackrel{\rm def}{=} PHP + V_{\parallel}(t),
\label{l7b}
\end{equation}
one finds from (\ref{l1}), (\ref{l1a}) and (\ref{B1})
\begin{equation}
V_{\parallel} (t) U_{\parallel} (t) =
- i \int_{0}^{\infty} K(t - \tau ) U_{\parallel} ( \tau ) d \tau
\stackrel{\rm def}{=} - iK \ast U_{\parallel} (t) . \label{B3}
\end{equation}
(Here the asterisk $\ast$ denotes the convolution: $f  \ast
g(t) = \int_{0}^{\infty}\, f(t - \tau ) g( \tau  ) \, d \tau$ ).
\linebreak Next, using this relation and a retarded Green's operator  
$G(t)$ for the equation (\ref{B1})
\begin{equation}
G(t) = - i {\mit \Theta} (t) \exp (-itPHP)P,
\label{B4}
\end{equation}
one obtains \cite{pra,9,10}
\begin{equation}
V_{\parallel}(t) \; U_{\parallel}(t) =
- i K \ast \Big[ {\it 1} + \sum_{n = 1}^{\infty} (-i)^{n}L
\ast \ldots \ast L \Big] \ast U_{\parallel}^{(0)} (t) ,
\label{B5}
\end{equation}
where $L$ is convoluted $n$ times, ${\it 1} \equiv
{\it 1}(t) \equiv \delta (t)$,
\begin{equation}
L(t) = G \ast K(t), \label{B6}
\end{equation}
and
\begin{equation}
U_{\parallel}^{(0)} = \exp (-itPHP) \; P \label{B7}
\end{equation}
is a "free" solution of Eq. (\ref{B1}). Of course,
the  series (\ref{B5}) is convergent if \linebreak
$\parallel L(t) \parallel < 1$. If for every $t \geq 0$
\begin{equation}
\parallel L(t) \parallel \ll 1, \label{B8}
\end{equation}
then, to the lowest order of  $L(t)$,  one  finds  from
(\ref{B5}) \cite{pra,9,10}
\begin{equation}
V_{\parallel}(t) \cong V_{\parallel}^{(1)} (t)
\stackrel{\rm def}{=} -i \int_{0}^{\infty} K(t - \tau )
\exp {[} i ( t - \tau ) PHP {]} d \tau , \; \; (t \geq 0). \label{B9}
\end{equation}

From (\ref{B3}) and (\ref{B9}) it follows that
\begin{equation}
V_{\parallel}(t = 0) \equiv V_{\parallel}^{(1)}(t = 0) = 0, 
\label{v-0}
\end{equation}
which  (by (\ref{l7b}) ) means that 
\begin{equation}
H_{\parallel}(t = 0) \equiv PHP , \label{h-0}
\end{equation}
and thus in the general case
\begin{equation}
h_{jk} (t = 0) \equiv H_{jk}, \; \; \; v_{jk} (t = 0 ) \equiv 0, 
\label{h-jk-0}
\end{equation}
where $v_{jk}(t) = <{\bf j}|V_{\parallel}(t)|{\bf k}>$, $j,k =1,2$.

In the case of (\ref{l3}) of the projector $P$, the approximate formula 
(\ref{B9}) for $V_{\parallel}(t)$ enables us to calculate the matrix 
elements $v_{jk}(t > 0)$ of $V_{\parallel}(t) 
\cong V_{\parallel}^{(1)}(t)$ \cite{9,10}, which leads to the following  
expressions for $v_{jk}(t \rightarrow \infty )\stackrel{\rm  def}{=}  
v_{jk}$ (for details see  \cite{9,10}),
\begin{eqnarray}
v_{j1} = & - & \frac{1}{2} \Big( 1 + \frac{H_{z}}{\kappa} \Big)
{\Sigma}_{j1} (H_{0} + \kappa ) -
\frac{1}{2} \Big( 1 - \frac{H_{z}}{\kappa} \Big)
{\Sigma}_{j1} (H_{0} - \kappa )  \nonumber \\
& - & \frac{H_{21}}{2 \kappa} {\Sigma}_{j2} (H_{0} + \kappa )
+ \frac{H_{21}}{2 \kappa} {\Sigma}_{j2} (H_{0} - \kappa ) , 
\nonumber \\
& \, & \label{B10} \\
v_{j2} = & - & \frac{1}{2} \Big( 1 - \frac{H_{z}}{\kappa} \Big)
{\Sigma}_{j2} (H_{0} + \kappa ) -
\frac{1}{2} \Big( 1 + \frac{H_{z}}{\kappa} \Big)
{\Sigma}_{j2} (H_{0} - \kappa ) \nonumber  \\
& - & \frac{H_{12}}{2 \kappa} {\Sigma}_{j1} (H_{0} + \kappa )
+ \frac{H_{12}}{2 \kappa} {\Sigma}_{j1} (H_{0} - \kappa ) , 
\nonumber
\end{eqnarray}   
where $j,k = 1,2$,
\begin{equation}
H_{z} = \frac{1}{2} ( H_{11} - H_{22} ) , \label{B11}
\end{equation}
and 
\begin{equation}
H_{0} =  \frac{1}{2} ( H_{11} + H_{22} ),
\label{b12},
\end{equation}
\begin{equation}
\kappa = ( |H_{12} |^{2} + H_{z}^{2} )^{1/2} . \label{B12}
\end{equation}
Hence, by (\ref{l7b})
\begin{equation}
h_{jk} = H_{jk} + v_{jk} , \label{B13}
\end{equation}
which defines the operator $H_{\parallel} \stackrel{\rm def}{=}
H_{\parallel} (t \rightarrow \infty )  \stackrel{\rm def}{=}
PHP + V_{\parallel}^{(1)}( \rightarrow \infty ) \equiv 
H_{\parallel}(\infty)$. It should be emphasized that all the components 
of the expressions (\ref{B10}) are of the same order with respect to 
$\Sigma (\epsilon )$. 

These formulae for  $v_{jk}$  and  thus  for $h_{jk}$  have  been
derived without assuming any symmetries of the type  CP--,  T--,  or
CPT--symmetry  for  the  total  Hamiltonian  H   of   the   system
considered. According to the general ideas of the quantum theory, one  
can state  that  the operator $H_{\parallel}(t \rightarrow \infty)
\equiv H_{\parallel}(\infty)$ describes the bounded or
quasistationary states  of  the  subsystem  considered. 

The same formula for $H_{\parallel}(\infty)$ can be obtained 
by improving the method described in \cite{1,2}. Namely, analysing the 
LOY derivation of the effective Hamiltonian for the neutral kaon complex 
one can observe that the components containing the matrix elements 
$<{\bf j}|H_{I}|{\bf k}>$, $(j,k =1,2)$ are neglected in the right 
sides of Eqs (18), (19) in \cite{1}.  It is found in \cite{improved} 
that such an approximation is not true for the whole domain of the 
parameter $t$: It is not true for $t \simeq t_{0} = 0$.  The formulae
for the improved effective Hamiltonian $H_{LOY}^{Imp}$ can be obtained
by considering the equations for the component 
$|\psi >_{\parallel}$ of the state vector $|\psi >$  
instead of the LOY equations for the amplitudes $a_{1}, a_{2}$ (see Eq 
(\ref{l1b}) ) mentioned above. This approach takes into account the 
matrix elementsof $H_{I}$ which were neglected in the LOY paper 
\cite{1}. It appears that  such $H_{LOY}^{Imp}$ 
equals $H_{\parallel}(\infty)$ exactly \cite{improved}.

Properties of the matrix elements $h_{jk}$ of the  apprroximate 
$H_{\parallel}(t)$ described in this Subsection have been examined in 
\cite{9} for the generalized Fridrichs--Lee model \cite{chiu}. 
For this model it has been found in \cite{9} that 
$h_{jk}(t) \simeq h_{jk}$ practically for $t \geq T_{as} 
\simeq \frac{10^{2}}{\pi (m_{0} - |m_{12}| - \mu )}$, 
where $m_{0} \equiv H_{11} = H_{22}, \, m_{12} \equiv H_{12}$ and 
$(m_{0} - \mu )$ is the the diffrence between the mass of the unstable 
particles considered and the threshold energy of the continuum state of 
decay products (see \cite{9}, formula (153)). For the neutral K--system, 
to estimate $T_{as}$ it has been taken $(m_{0} - |m_{12}| - \mu ) \simeq 
(m_{0} - \mu ) = m_{K} - 2m_{\pi} \sim 200$MeV, which gives 
$T_{as} \sim 10^{-22}$ sec \cite{is}.

\section{Consequences of CPT--invariance.}

In the case of preserved  CPT symmetry,
\begin{equation}
\Theta  H {\Theta}^{-1} = H, \label{l19}
\end{equation}
and violated CP,
\begin{equation}
[{\cal C}{\cal P},H] \neq 0, \label{cp}
\end{equation}
assuming (\ref{cpt1})  one finds 
\begin{equation}
H_{11} = H_{22},  \label{H11=H22}
\end{equation}
which implies that for $t = 0$ (see (\ref{h-jk-0}) )
\begin{equation}
h_{11}(0) \equiv h_{22}(0) .  \label{h11=h22}
\end{equation} 
This property means that 
\begin{equation}
\Re \,( h_{11}(0) ) = M_{11}(0) \equiv \Re \, (h_{22}(0))  = M_{22}(0),
\label{m11=m22}
\end{equation}
i.e., in CPT--invariant system a particle and its antiparticle are 
created at $t = 0$ as quantum objects with equal masses.

From (\ref{H11=H22}) it also follows that $\kappa \equiv |H_{12} |$, 
$H_{z} \equiv 0$ and  $H_{0}\equiv H_{11} \equiv H_{22}$, and 
\cite{9,10}
\begin{equation}
{\Sigma}_{11} ( \varepsilon = {\varepsilon}^{\ast} ) \equiv
{\Sigma}_{22} ( \varepsilon = {\varepsilon}^{\ast} )
\stackrel{\rm def}{=} {\Sigma}_{0} ( \varepsilon
= {\varepsilon}^{\ast} ) . \label{B14}
\end{equation}
Therefore for $t \rightarrow \infty $ the matrix  elements  
$v_{jk}^{\mit \Theta}$  (\ref{B10})  of operator 
$V_{\parallel}^{\mit \Theta}$ ($V_{\parallel}^{\mit \Theta}$ denotes 
$V_{\parallel}$ when (\ref{l19}) occurs) take the following form
\begin{eqnarray}
v_{j1}^{\mit \Theta} = & - & \frac{1}{2} {\Big\{ } {\Sigma}_{j1} 
(H_{0} + |H_{12} |)
+ {\Sigma}_{j1} (H_{0} - | H_{12} |) \nonumber  \\
& + & \frac{H_{21}}{|H_{12}|} {\Sigma}_{j2} (H_{0} + | H_{12} |)
- \frac{H_{21}}{|H_{12}|} {\Sigma}_{j2} (H_{0} - | H_{12} |) 
{\Big\} } , \nonumber \\
& \, & \label{B15} \\
v_{j2}^{\mit \Theta} = & - & \frac{1}{2} {\Big\{ } {\Sigma}_{j2} 
(H_{0} + |H_{12} |)
+ {\Sigma}_{j2} (H_{0} - | H_{12} |)  \nonumber \\
& + & \frac{H_{12}}{|H_{12}|} {\Sigma}_{j1} (H_{0} + | H_{12} |)
- \frac{H_{12}}{|H_{12}|} {\Sigma}_{j1} (H_{0} - | H_{12} |) 
{\Big\} }.  \nonumber
\end{eqnarray}

Using these relations one easily finds 
\begin{eqnarray}
h_{11}^{\mit \Theta} - h_{22}^{\mit \Theta}  = & - &
\frac{1}{2|H_{12}|} \Big[
H_{21} \, {\Sigma}_{12} (H_{0} + |H_{12}|) -
H_{12} \, {\Sigma}_{21} (H_{0} + |H_{12}|) \nonumber \\
& - & 
H_{21} \, {\Sigma}_{12} (H_{0} - |H_{12}|) +
H_{12} \, {\Sigma}_{21} (H_{0} - |H_{12}|) \Big], \label{h11-h22}
\end{eqnarray}
where $h_{jk}^{\mit \Theta}$ is defined analogously to 
$v_{jk}^{\mit \Theta}$.  From this Equation it follows that within 
the more accurate approximation than the one used by Lee, Oehme and 
Yang there is
\begin{equation}
h_{11}^{\mit \Theta} - h_{22}^{\mit \Theta} \neq 0. \label{h11-h22a}
\end{equation}
So, the more exact approximation does not confirm the standard LOY 
result (\ref{b8}). Expression (\ref{h11-h22}) also impairs a conviction 
that a particle and its antiparticle have always the same mass in 
CPT--invariant system, i.e., the conclusion (\ref{m-loy}) of the LOY 
theory. Namely, taking into account the decomposition (\ref{r24}) it is 
not difficult to obtain the following relation from (\ref{h11-h22}) 
\begin{eqnarray}
\Re \, (h_{11}^{\mit \Theta} - h_{22}^{\mit \Theta}) & = &
\Im \, \Big\{ \frac{H_{21}}{|H_{12}|} \, 
{\Sigma}_{12}^{I} (H_{0} + |H_{12}|) \Big\} \nonumber \\
& - & \Im \, \Big\{ \frac{H_{21}}{|H_{12}|} \, 
{\Sigma}_{12}^{I} (H_{0} - |H_{12}|) \Big\}   
\label{m11-m22} \\
& \stackrel{\rm def}{=} & M_{11}  -  M_{22} \equiv \Delta m, 
\nonumber
\end{eqnarray}
which means that the masses of a particle and its simultaneously 
prepared antiparticle need not be equal at $t \rightarrow \infty$ if  
CPT--symmetry holds. Only at $t = 0$, i.e., at the moment of the 
creation of a particle and its antiparticle  their masses can be equal 
(\ref{m11=m22}). The sign of $\Delta m$ depends on the form of $H$.

It can be expected that similar conclusions can be drawn using the 
approach exploited in \cite{3}, where perturbation theory 
correction improving \linebreak $H_{LOY}$ have been considered.

\section{Possible experimental consequences.}

It seems that the only way for an experimantal verification of the
effect expressed by formula (\ref{m11-m22}) in the near future
is to search for the 
properties of $K_{0}, \overline{K_{0}}$ and similar complexes.
The real part of $(h_{11} - h_{22})$ can be expressed by the parameters
measured in the experiments with neutral $K$, or $B$ mesons \cite{4}
--- \cite{dafne}, \cite{data}, that is, it is an experimentally 
measurable quantity. Within the use of the formalism described in Sec. 
2.2, $(h_{11} - h_{22})$ has been estimated for the generalized 
Fridrichs--Lee model \cite{chiu}. Assuming CPT--invariance (i.e., 
(\ref{l19}) ) and $|m_{12}|\equiv |H_{12}| \ll (m_{0}- \mu ) \equiv 
(H_{0} - \mu )$ it  has been  found  in  \cite{is} that 
\begin{eqnarray}
\Re \, (h_{11}^{FL}  -  h_{22}^{FL}) 
& \simeq &
i \frac{ m_{21}{\Gamma}_{12} - m_{12}{\Gamma}_{21} }{4(m_{0} - \mu )}
\nonumber \\
& \equiv & \frac{ {\Im} \, (m_{12}{\Gamma}_{21}) }{2(m_{0}- \mu )},
\label{FL2}  
\end{eqnarray}
where $h_{jj}^{FL}, \; (j =1,2),$ denotes $h_{jj}$ calculated for the 
Fridrichs--Lee model. For the ${K_{0}}, \overline{K_{0}}$--complex 
${\Gamma}_{21} \equiv {\Gamma}_{12}^{\ast} \simeq \Re \,
({\Gamma}_{12})
\simeq \frac{1}{2}{\Gamma}_{s} \sim 3,7 \times 10^{-12}$MeV. This 
property and relation (\ref{FL2}) enable us to find the following 
estimation of $\Re \, (h_{11}^{FL}  -  h_{22}^{FL})$ for the neutral 
K--system \cite{is}
\begin{equation}
\Re \, (h_{11}^{FL}  -  h_{22}^{FL}) 
\sim 9,25 \times 10^{-15} \Im \, (m_{12})
\equiv 9,25 \times 10^{-15} \Im \, (H_{12}) . \label{FL}
\end{equation}
The assumptions leading to (\ref{FL2}) can be used to obtain
\begin{equation}
h_{11}^{FL} + h_{22}^{FL}  \simeq  2m_{0} - i {\Gamma}_{0} -
\frac{i}{2} \frac{ \Re \,(H_{21} {\Gamma}_{12} )}{m_{0} - \mu},
\label{2h-0}
\end{equation}
(where ${\Gamma}_{0} \equiv {\Gamma}_{11} = {\Gamma}_{22}$ ) which 
and (\ref{FL2}), (\ref{FL}) give 
\begin{equation}
\frac{\Re \, (h_{11}^{FL}  -  h_{22}^{FL}) }
{\Re \, (h_{11}^{FL}  +  h_{22}^{FL}) } \simeq
\frac{ \Im \, (H_{12})}{8m_{0}(m_{0} - \mu )} {\Gamma}_{s} 
\sim 9,25 \times 10^{-18} \, (\Im \, (H_{12})) \; [{\rm MeV}]^{-1},
\label{hz-h0}
\end{equation}
(where  $m_{0} = m_{K}$ is inserted). This estimation 
does not contradict the 
experimental data for neutral K mesons \cite{data}. 

From (\ref{FL}) it follows that the effect described 
by relation (\ref{m11-m22}) is very, very small indeed, 
and it is, probably, 
beyond the accuracy of today's experiments \cite{data}. Test of more 
higher accuarcy are expected to be performed in the near future 
\cite{dafne}. Advances in the experimental methods are planned in 
DA$\Phi$NE project at Frascati (Rome) and BARBAR at SLAC (Stanford) to 
measure the smallobservables for neutral meson systems. So, there is a 
chance that the effects described in Sec. 3 can be confirmed 
experimentally in the near future.
  
Using  the formalism described in Sec. 2.2 it has been shown in 
\cite{9,10,urb-pla} that eigenvectors
$|l^{t}>, |s^{t}>$,
\begin{eqnarray}
|l^{t}> & = \frac{1}{(1 + | {\alpha}_{l} (t) |^{2})^{1/2}}
[|{\bf 1}> - {\alpha}_{l}(t) |{\bf 2}>], \label{l-s} \\
|s^{t}> & = \frac{1}{(1 + | {\alpha}_{s} (t) |^{2})^{1/2}}
[|{\bf 1}> - {\alpha}_{s}(t) |{\bf 2}>], \nonumber
\end{eqnarray}
(for the definitions of ${\alpha}_{l}(t), {\alpha}_{s}(t)$ see 
\cite{9} )
of the more accurate effective Hamiltonian, 
$H_{\parallel}(t)$, for the neutral $K$ mesons complex possess the 
property $|l(s)^{t=0}> \neq |l(s)^{t \rightarrow \infty}>
\stackrel{\rm def}{=} |l(s)>$ if the CP symmetry is violated (i.e., 
if (\ref{cp}) holds). This property is the consequence of the fact 
that $H_{\parallel}(t=0)$ $\neq H_{\parallel}(t \rightarrow \infty)$. 
For the eigenstates $|K_{L(S)}>$ of $H_{LOY}$ such an effect is 
absent. (The eigenstates $|l^{t}>, |s^{t}>$ for $H_{\parallel}(t)$   
correspond to the eigenvectors $|K_{L(S)}>$ for $H_{LOY}$). 
The property of $|l^{t}>, |s^{t}>$ mentioned above means that the real 
properties of the system created at $t_{0}=0$ should be different
at $t=t_{0}=0$ and at $t \gg t_{0}$, i.e., at $t \rightarrow \infty$.
Relations  (\ref{h11=h22}), (\ref{m11=m22}) and (\ref{h11-h22a}), 
(\ref{m11-m22}) are a consequnence of this effect. 
Another implication of the above is that 
$|<l(s)^{t=0}|l(s)>| \neq 1$, 
although $<l(s)^{t=0}|l(s)^{t=0}> =<l(s)|l(s)> = 1$. 
The possibility to verify  of this effect by an experiment has been 
discussed in \cite{9,10,urb-pla}. The following problem arises: The 
states $|l^{t>0}>,|s^{t>0}>$ are not orthogonal, $<l^{t>0}|s^{t>0}> 
\neq 0$, and similarly ${<t; l^{t}|s^{t} ; t>|_{t >  T_{as}} }
\equiv$ ${<t; l|s; t>|_{t > T_{as}} }\neq 0$, where  
$|l(s); t>|_{t >T_{as}} \simeq \exp {(-itH_{\parallel})}  
|l(s)>|_{t > T_{as} }  $. 
This means that at time $t$ an observer is unable to register 
separatelly decay products of the state $|s^{t}>$ and of $|l^{t}>$ 
\cite{10}.  Nevertheless, 
it seems that this difficulty can be overcome. Namely, taking 
into account that for the neutral $K$ mesons system ${\tau}_{l(s)} 
\gg T_{as}$
and ${\tau}_{l} \simeq 0,58 \times 10^{2} {\tau}_{s}$, \cite{data}, 
one finds 
\linebreak
${|<t; l|s ; t>|^{2}_{t \sim {\tau}_{s}} }$ 
$\sim  e^{-1} \; {|< l|s>|^{2}} \simeq e^{-1} \;
{|< K_{L}|K_{S}>|^{2} }$ and
\linebreak[4]
$|<t; l|s ; t>|^{2}_{t \sim {\tau}_{l}}$
$\sim  e^{-1} \; (e^{-0.58})^{100} \, |< l|s>|^{2} \simeq 
e^{-1} (e^{-0.58})^{100} \, {|< K_{L}|K_{S}>|^{2}}$ 
\linebreak[4]
$\cong 0$. This last
estimation menas that for times $ t \sim {\tau}_{l}$ and 
$t > {\tau}_{l}$ the result of the observation of the decay process 
of states $|l>$ is practically not disturbed by the decay of states 
$|s>$. Now we can find the probability $p_{l}(t)$ of finding the system
in the particle state of type $|l^{t}>$ at a relatively long time, 
$t \gg T_{as}$, if it were in the state of this same type at $t=0$, 
(i.e., in $|l^{t=0}>$ ). It equals \cite{9,urb-pla}
\begin{eqnarray}
p_{l}(t \gg T_{as}) & = & 
\simeq  |<l^{t=0}|e^{-i t H_{\parallel} }|l>|^{2} 
{\mid}_{t \gg T_{as} } \nonumber \\
& = & {\rm exp}(- t {\Gamma}_{l} ) |<l^{t=0}|l>|^{2} . \label{p-l}
\end{eqnarray}
From this, the conclusion that there is a chance of observing a new
effect which is absent in the standard LOY theory was drawn in 
\cite{urb-pla}. Simply, within the LOY approach one has 
\begin{eqnarray}
p_{l}(t) \equiv
p_{K_{L}}(t) &=& |<K_{L}^{t=0}|e^{\textstyle (-itH_{LOY})}
|K_{L}>|^{2}  \nonumber \\
& \equiv &  |<K_{L}|e^{\textstyle (-itH_{LOY})}
|K_{L}>|^{2}  \simeq {\rm exp}(- t {\Gamma}_{l}) .
\label{p-KL} 
\end{eqnarray}
Now, if one has the possibility of determing the function $p_{l}(t)$ 
directly from the experiment, then starting from this experimentally 
determined $p_{l}(t)$ one can draw a segment of a straight line
for $t \sim {\tau}_{l}$ and $t > {\tau}_{l}$, 
\begin{equation}
y_{l}(t) 
\stackrel{\rm def}{=} \ln  p_{l}(t) + b_{l},
\; \; \; (t \sim {\tau}_{l}, t > {\tau}_{l}),
\label{y-t}
\end{equation}
where
\begin{equation}
b_{l} = \ln |<l^{t=0}|l>|^{2} .
\label{b-l}
\end{equation}
(The time $t=0$ is the instant of the creation of the neutral $K$ 
meson). Then one can produce this segment extrapolating it to the 
intersection with the $y$--axis. The point $y(0)$ determines the 
value of the parameter $b_{l}$. The ap\-pro\-xi\-ma\-tion 
described in Sec. 2.2 (which is more accurate than the LOY method)
predicts that there should be  $b_{l} \neq 0$ in 
the CPT invariant but CP noninvariant system, whereas within 
the LOY theory one finds $b_{l}^{LOY} = 0$.  This situation is 
presented in Fig. 1. In this case $\tan {\varphi} = {\Gamma}_{l}$. \\
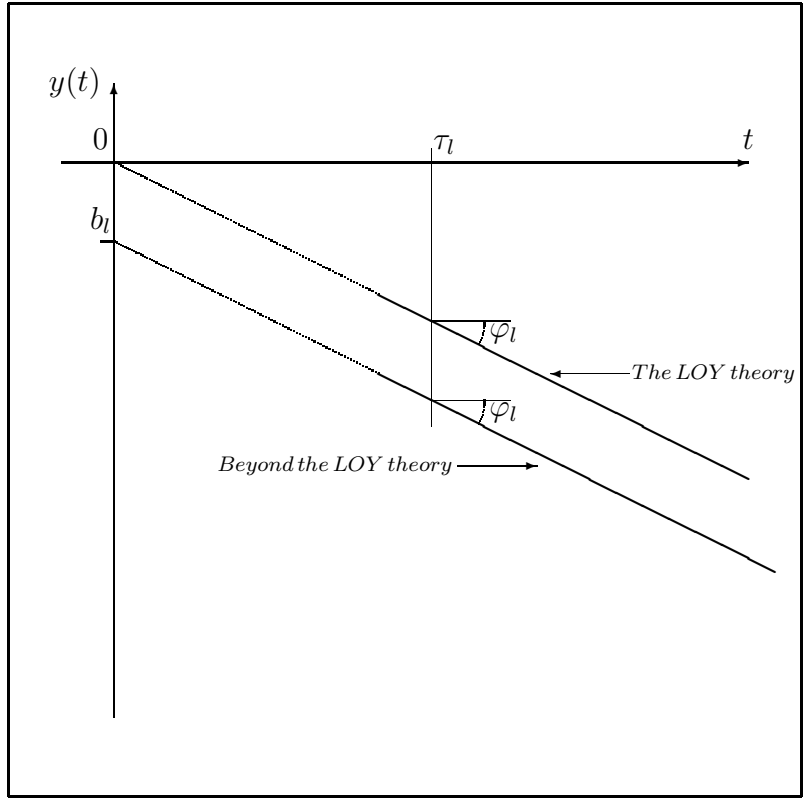
\begin{figure}[h]
\centering
\begin{picture}(350,350)(-20,0)
{\linethickness{0.6pt}
\put(0,300){\line(0,-1){300}}
\put(0,300){\line(1,0){300}}
\put(300,300){\line(0,-1){300}}
\put(40,30){\vector(0,1){240}}
\put(20,240){\vector(1,0){260}}
\put(300,0){\line(-1,0){300}}  }
{\linethickness{0.1pt}
\put(160,245){\line(0,-1){105}} }
\put(35,270){\makebox(0,0)[r]{$y(t)$}}
\put(280,245){\makebox(0,0)[b]{$t$}}
\put(165,245){\makebox(0,0)[b]{${\tau}_{l}$}}
\put(35,245){\makebox(0,0)[b]{$0$}}
\bezier{100}(40,240)(80,220)(140,190)
\bezier{100}(40,210)(80,190)(140,160)
\put(40,210){\line(-1,0){5}}
\put(35,215){\makebox(0,0)[b]{$b_{l}$}}
\put(140,190){\thicklines {\line(2,-1){140}}}
{\linethickness{0.1pt}
\put(160,180){\line(1,0){30}}}
\bezier{10}(180,180)(180,175.279)(177.889,171.056)
\put(182,175.279){\makebox(0,0)[l]{${\varphi}_{l}$}}
{\linethickness{0.1pt}
\put(160,150){\line(1,0){30}}}
\bezier{10}(180,150)(180,145.279)(177.889,141.056)
\put(182,145.279){\makebox(0,0)[l]{${\varphi}_{l}$}}
\put(140,160){\thicklines {\line(2,-1){150}} }
{\linethickness{0.1pt}
\put(235,160){\vector(-1,0){30}} }
\put(236,160){\makebox(0,0)[l]{{$\scriptstyle The \,LOY \, theory$}} }
{\linethickness{0.1pt}
\put(170,125){\vector(1,0){30}}}
\put(168,125){\makebox(0,0)[r]{{$\scriptstyle Beyond \, the \, LOY \, 
theory$}}}
\end{picture}
\caption{The hypothetical form of the extrapolated 
experimentally obtained (bold line) curve $y_{l}(t) = \ln  p_{l}(t) = 
- {\Gamma}_{l} t + b_{l}$.  }
\end{figure}

The following estimation has been found for the value of 
the parameter $b_{l}$ in \cite{urb-pla}: $b_{l} \simeq -2.67 \times
10^{-6}$. 

It can be easily shown that if the property (\ref{l19}) holds then 
$H_{\parallel} =  H_{LOY}$ if and only if $|H_{12}| = 0$, and 
similarly, following \cite{is}, that  $(h_{11} - h_{22}) = 0$ if 
and only if
$|H_{12}| = 0 $. This means that if $|H_{12}| \equiv 
|<{\bf 1}|H_{I}|{\bf 2}>| \neq 0$ then there should be $H_{\parallel} 
\neq H_{LOY}$ and thus $b_{l} \neq 0$. 

Any experimental confirmation of this effect, i.e, that $b_{l} \neq 0$, 
will mean,  contrary to the predictions of the LOY theory,  
that if CP symmetry is violated then the properties of the system at 
the instant of its creation, $t = t_{0} \equiv 0$,  and at time
$ t \gg t_{0}$ ($ t \rightarrow \infty$) are different.  
Indirectly, after verifying  if $H_{12} \neq 0 $,
it will also mean  that the other conclusions derived
within the use the more accurate approximation described in Sec. 2.2,
(including results obtained in Sec. 3)  should be true.

There is another relation which seems to be useful
when one tries to verify 
predictions based on the aproximation described in Sec. 2.2. Namely, 
one can find that
\begin{equation} 
{\alpha}_{l} + {\alpha}_{s} = \frac{h_{11} - h_{22}}{h_{12}},
\label{al-as}
\end{equation}
(see \cite{9,10}). Within the LOY theory one simply has 
${\alpha}_{l}^{LOY} \stackrel{\rm def}{=} {\alpha}^{LOY}  = - 
{\alpha}_{s}^{LOY}$, that is, ${\alpha}_{l}^{LOY} + 
{\alpha}_{s}^{LOY} = 0$. So, if $(h_{11} - h_{22}) \neq 0$ then 
measurements determining the values of the parameters ${\alpha}_{l}$ 
and ${\alpha}_{s}$ more accurately than it is possible in todays 
experiments should confirm this property.

\section{Discussion.}
\subsection{General remarks.}

The real parts of the diagonal matrix elements of the mass matrix 
$H_{\parallel}$, $h_{11}$ and $h_{22}$, are considered in the 
literature as masses of particles $|{\bf 1}>, |{\bf 2}>$ (eg., mesons 
${\rm K}_{0}$ and ${\overline{\rm K}}_{0}$, etc.,) \cite{1} --- 
\cite{baldo}. Result (\ref{m11-m22}) means that if 
${\Sigma}_{jk}^{I}(H_{0} + |H_{12}|) \neq 0$, or if 
${\Sigma}_{jk}^{I}(H_{0} - |H_{12}|) \neq 0$, ($j,k = 1,2$), or if 
both these cases occur, i.e., if particle "1" and antiparticle "2" are 
unstable and if they were simultneously prepared at $t = t_{0} 
\equiv 0$, then at $t \gg t_{0}$ (in particular at
$t \rightarrow \infty$) the masses of a decaying particle "1" and its 
antiparticle "2" should be different if CPT--symmetry is conserved and 
CP is violated in the system containig these unstable particles. In 
other words, unstable states $|{\bf 1}>, |{\bf 2}>$ appear to be 
nondegenerate in mass if CPT--symmetry holds and CP does not in the 
total system considered.  

On the other hand, examining properties of the operator 
${\Sigma}_{jk}^{I}( \epsilon )$ (\ref{r24b}) one finds that
\begin{equation}
{\Sigma}_{jk}^{I}(H_{0} \pm |H_{12}|) =  0 \; {\rm if} \;
(H_{0} \pm |H_{12}|) < {\varepsilon}_{M}, \label{e-m}
\end{equation}
where ${\varepsilon}_{M}$ denotes the lower bound for the continous 
part ${\sigma}_{c}(QHQ)$ of the spectrum ${\sigma}(QHQ)$ of the 
operator $QHQ$. States $|{\bf 1}>, |{\bf 2}>$, for which this property 
occur, correspond to bound states and they cannot decay at all. This 
observation and relation (\ref{m11-m22}) mean that in the 
CPT--invariant system the masses of a given particle and its aniparticle 
prepared at the finite instant $t = t_{0} > - \infty$ are equal (i.e., 
apear to be degenerate) only in the case of bound (stable) states 
$|{\bf 1}>, |{\bf 2}>$. The case when vectors $|{\bf 1}>, |{\bf 2}>$ 
describe pairs of particles $p, \overline{p}$, or $e^{-}, e^{+}$, can 
be considered as an example of such states. 

In fact there is nothing strange in these conlusions. From (\ref{l19}) 
(or from the CPT Theorem \cite{cpt} ) it only follows that the masses of 
particle and antiparticle eigenstates for $H$ (i.e., masses of 
stationary states of $H$) should be the same in CPT invariant system. 
Such a conclusion can not be derived from (\ref{l19}) for particle 
$|{\bf 1}>$ and its antiparticle $|{\bf 2}>$ if they are unstable, 
i.e., if states $|{\bf 1}>, |{\bf 2}>$ are not eigenstates of $H$. 
One should remember that the CPT Theorem of axiomatic quantum field 
theory has been proved for quantum fields corresponding to stable 
quantum objects. Only such fields are considered in axiomatic quantum 
field theory \cite{cpt}. There is no axiomatic quantum field theory of 
unstable quantum particles. So, all implications of CPT Theorem 
(including those obtained within the S--matrix method) need not be valid 
for decaying particles prepared at some initial instant $t_{0} = 0$ 
and then evolving in time $t \geq 0$. Simply, the consequences of CPT 
invariance need not be the same for systems in which time $t$ varies 
from $t = - \infty$ to $t = + \infty$ and for the system in which $t$ 
can vary only from $t = t_{0} > - \infty$ to $t = + \infty$.

The following conclusion can be drawn from (\ref{m11=m22}) and 
(\ref{m11-m22}): The time evolution  causes 
the masses of the unstable particle 
and its simultaneously prepared at $t=0$ antiparicle 
to be different at 
$t >  T_{as}$ in the CPT invariant but CP noninvariant system.

All the above conclusions contradict the standard result of the LOY 
and related approaches. Properties of the real systems described by 
formulae (\ref{h11-h22}), (\ref{m11-m22}) are unobservable for the LOY 
approximation. On the other hand, these relations are not in conflict 
with the general conclusions of \cite{11,14}, where CPT--transformation 
properties of the exact $H_{\parallel}(t)$ have been studied. 
Confronting relation (\ref{b8}) with (\ref{h11-h22}) one should remember 
that, in fact, the LOY model is unable to describe real properties of 
the system considered \cite{is,why,kabir}. Namely, it has been proved 
in \cite{is,why} that systems containing an exponentially decaying 
subsystem (i.e., evolving in time according to (\ref{form})) cannot be 
CPT--invariant. A similar conclusion can be drawn from the results 
obtained in \cite{kabir}. This means that the CPT--symmetry properties 
of the LOY model need not reflect real properties of the system 
considered.  At the same time from \cite{11,14} it follows that CPT-- 
and other transformation properties of the effective Hamiltonian 
described in Sec. 2.2 are consistent with properties of the real 
systems.

Note that considering in detail the generalized Fridrichs--Lee model 
one finds that ${\Gamma}_{jk} = 0, \; (j,k =1,2)$ for $m_{0} < 
\mu$, i.e., for bound states \cite{chiu, 9}. This observation and 
relation (\ref{FL2}) imply that for bound (stable) states $\Re \, 
(h_{11}^{FL} - h_{22}^{FL} )= 0$. So, if CPT--symetry is conserved in 
this model, then particle and antiparticle bound states remain to be 
also degenerate in mass beyond the LOY approximation, whereas unstable 
states (i.e., states for which $m_{0} > \mu$) appear to be 
nondegenerate in mass in this model if CPT--symetry holds but CP does 
not. These observations confirm our earlier conlusions following from 
(\ref{m11-m22}), (\ref{e-m}). 

As it was mentioned in Sec. 4,  one can conclude from (\ref{FL}), 
(\ref{hz-h0}) that the relation (\ref{m11-m22}) leads to a very 
small value of ${\Delta}m$, which is probably, for single 
"particle--aniparticle" pair, beyond today's experiments accuracy.
Nevertheless, this effect should be significant 
if the number of pairs of unstable particles and their antiparticles 
is very large. Such conditions can be met at the origin immediately 
after the "Big Bang", at the first instants of the existence of our 
Universe \cite{m-antim} --- \cite{m-antim1a}. (Note, that the initial 
condition (\ref{init}) corresponds exactly to the case of the creation 
of the Universe). 

\subsection{Possible cosmological applications.}

In \cite{m-antim11} --- \cite{m-antim1a} an observation has been 
used that a baryon asymmetry could arise even in thermal equilibrium if
CPT symmetry is violated. It is assumed in such theories that CPT
is not realized as a good symmetry in the early Universe. 
From the main result of Sec. 3,  (\ref{m11-m22}), it follows that
in fact CPT need not be violated in order that the baryon asymmetry 
could occur. Indeed, the termodynamic observable number density, 
$n_{X}^{eq,\pm}$, in thermal equilibrium, is a function of 
temperature alone:
\begin{equation}
n_{X}^{eq,\pm} (T)= g^{s}_{X} \int \frac{d^{3} \vec{p}}{(2 \pi )^{3}} 
f^{\pm}_{X}(\vec{p} ,T),
\label{N-eq}
\end{equation}
where $g^{s}_{X}$ is the number of spin states of particle type $X$,
\begin{equation}
f^{\pm}_{X}(\vec{p} ,T) = 
\frac{1}{e^{\textstyle (E_{X} - {\mu}_{X} )/T} \pm 1},
\label{n-eq-x}
\end{equation}
is the equilibrium phase space occupancy,
${\mu}_{X}$ is a possible chemical potential of the particles 
considered,
$E_{X} = \sqrt{m^{2} + p^{2}}$ and $\vec{p}$ is the momentum,
$p = |\vec{p}|$, $\pm$ refers to Bose--Eistein ($-$), and Fermi--Dirac
($+$) statistics \cite{m-antim} --- \cite{m-antim1a}. 

In the literature
the relation 
\begin{equation}
n^{eq,+}_{X}(T) = n^{eq,+}_{{\overline X}}(T),  \label{n-eq}
\end{equation}
which can be obtained from (\ref{n-eq-x}) assuming 
that masses, $m$, of particle, $X$,  and $\overline{m}$ of 
antiparticle, ${\overline X}$, are equal, (i.e., that $E_{X} = 
E_{\overline X}$), and that 
baryon number $B$ is not conserved, (which, in thermal equilibrium,  
imlpies ${\mu}_{X} = 0$ \cite{m-antim} --- \cite{m-antim1a}),
is considered as the suggestion that without 
some extra mechanisms (e.g., a violation of CPT) 
the particle--antiparticle asymmetry can not be 
produced. 

Taking into account the results of Sec. 3 one finds that in fact there 
is $m \equiv m_{t}$, and 
\begin{equation}
m_{t=0} = {\overline{m}}_{t=0}, \label{mt0=mt0}
\end{equation}
but 
\begin{equation}
m_{t>T_{as}} \stackrel{\rm def}{=} m \neq {\overline{m}}_{t>T_{as}}
\stackrel{\rm def}{=} \overline{m}, \label{mT>mT}
\end{equation}
and thus $E_{X}  \equiv \sqrt{ m^{2} + p^{2} }
\neq E_{\overline X} \equiv \sqrt{{\overline{m}}^{2} + p^{2} }$ in 
the case of unstable particles $X$, which implies $n^{eq,\pm}_{X}(T) 
\neq n^{eq,\pm}_{{\overline X}}(T)$ instead of (\ref{n-eq}). 
The relation $n^{eq,\pm}_{X}(T) \neq n^{eq,\pm}_{{\overline X}}(T)$ 
means that  the asymmetric number of particles $X$,
and antiparticles ${\overline X}$, can be 
generated  at time $t > T_{as} \gg t = t_{0} =0$ 
in a CPT invariant but CP noninvariant system even when there is 
no the extra mechanism in this system and even  were the symetric
numbers, ${\cal N}^{X}_{t}$ of $X$, and ${\cal N}^{\overline{X}}_{t}$ 
of $\overline{X}$ at time $t = t_{0}=0$:
\begin{equation}
{\cal N}^{X}_{t=0} = {\cal N}^{\overline{X}}_{t=0}.
\label{N=N} 
\end{equation}
If to take into account that in the observed Universe
${\cal N}^{X}_{t} \gg {\cal N}^{\overline{X}}_{t}$, the following 
conclusion seems to be obvious
\begin{equation}
m_{t > 0} \, < \, {\overline{m}}_{t > 0}.
\label{mt>mt}
\end{equation}

Now let us examine 
the supposition that asymmetric numbers of particles and antiparticles
can be generated at time $t > T_{as}$ in thermal equilibrium.
Only the simplest
nontrivial example will be considered in this paper.
From the results of Sec. 3 it follows that stable 
particles cannot produce any contribution into the diference of masses 
particles and antiparticles but
only unstable particles can generate the matter--antimatter masses 
asymmetry. So, we will consider only  unstable $X$.
Assuming that the baryon number $B$ is not conserved we will take 
${\mu}_{X} = 0$  \cite{m-antim} --- \cite{m-antim1a}, (In general, for 
most purposes it is reasonable to set $\mu$ to be zero \cite{Sarkar}).
Thus the number density, $n^{eq,\pm}_{X}$, of unstable particles  $X$ 
in thermal equilibrium state equals
\begin{eqnarray}
n^{eq,\pm}_{X}(T) & = & g^{s}_{X} 
\int \frac{d^{3}\vec{p}}{(2 \pi )^{3}} 
\frac{1}{e^{\textstyle  \frac{E_{X}}{T}} \pm 1} \nonumber \\
& = & 
\frac{2 g^{s}_{X}}{(2 \pi )^{2}}
\int_{m}^{\infty} 
\frac{E_{X} 
\sqrt{E_{X}^{2}-m^{2}}}{e^{\textstyle \frac{E_{X}}{T} }\pm 1}
dE_{X} \nonumber \\
& \equiv & 
\frac{ g^{s}_{X}}{2 {\pi}^{2}}
T^{3} I_{N}^{\pm}(a),
\label{N-eq-1}
\end{eqnarray}
where $a = \frac{m}{T}\, >\, 0$, and
\begin{equation}
I_{N}^{\pm}(a)  \stackrel{\rm def}{=} 
\int_{a}^{\infty}
\frac{x \sqrt{x^{2} - a^{2}}}{e^{\textstyle x} \pm 1} dx. \label{i-0a1}
\end{equation}

Similarly, for the number density $n^{eq,\pm}_{{\overline X}}$ of 
antiparicles ${\overline X}$ one finds
\begin{equation}
n^{eq,\pm}_{{\overline X}} (T)=
\frac{ g^{s}_{X}}{ 2 {\pi }^{2}} T^{3} I_{N}^{\pm}({\overline a}),
\label{N-eq-2b}
\end{equation}
where ${\overline a} = \frac{{\overline m}}{T}$.

From the observations of our Universe it follows that 
$n^{eq,\pm}_{X} (T) \, > \, n^{eq,\pm}_{{\overline X}} (T)$ at
time $t \gg T_{as}$. It can occur in thermal equilibrium only if $m <
\overline{m}$. So tere should be $\overline{m} = m + \Delta m$, and
$\Delta m > 0$. 
Now if we take into account the estimation (\ref{hz-h0}), from which it 
follows that $\Delta m$ should be much, much smaller than  $m$ and
${\overline m}$, and  define
\[
\sigma \stackrel{\rm def}{=} \frac{ \Delta m}{T} ,
\]
then we find for $\Delta m \ll  m$, i.e., for 
$a \equiv \frac{m}{T}$, ${\overline a} = \frac{{\overline m}}{T}$ and
$\sigma \ll a$
\begin{equation}
I_{N}^{\pm}({\overline a}) \equiv 
I_{N}^{\pm}(a + \sigma )
\simeq I_{N}^{\pm}(a) + {\sigma} I_{\Delta N}^{\pm}( a),
\label{i-0-1}
\end{equation}
where 
\begin{equation}
I_{\Delta N}^{\pm}(z)  \stackrel{\rm def}{=} 
{\frac{\partial I_{N}^{\pm}(x)}{\partial x} \, \vrule \,}_{x=z} .
\label{i-1-1}
\end{equation}
This means that the difference between the number 
densities of particles and 
antiparticles in the thermal equilibrium should be approximately equal 
\begin{equation}
n^{eq,\pm}_{X}(T) - n^{eq,\pm}_{{\overline X}}(T)
\simeq 
- \frac{ g^{s}_{X}}{2 {\pi}^{2}} T^{3} 
\Big( \frac{\Delta m}{T} \Big) {I_{\Delta N}^{\pm}(a)\, 
\vrule \,}_{a= \frac{m}{T}} .
\label{N-N}
\end{equation}
Using this relation the following ratio can be found
\begin{eqnarray}
{\eta}_{x}^{\pm}   \stackrel{\rm def}{=} 
\frac{n^{eq,\pm}_{X}(T) - n^{eq,\pm}_{{\overline X}}(T)}{n^{eq, 
\pm}_{X}(T)} & = & - \frac{\Delta m}{T} \, 
{ \frac{I_{\Delta N}^{\pm}(a)}{I_{N}^{\pm}(a)} \, 
\vrule \,}_{a = \frac{m}{T}}
\label{Lambda}\\
& \equiv & - \frac{\Delta m}{T} \, 
\Big\{
{ \frac{ \partial}{\partial z}  \ln I_{N}^{\pm}(z)
{\Big\}} \, \vrule \, }_{z= \frac{m}{T}} . \nonumber
\end{eqnarray}

Taking into accout the result (\ref{i-1-1=}) the
difference (\ref{N-N})  between number densities takes the form 
\begin{equation}
n^{eq,\pm}_{X}(T) - n^{eq,\pm}_{{\overline X}}(T)
\simeq  \mp
\frac{ g^{s}_{X}}{2 {\pi}^{2}} m^{2} \, \Delta m \,
\sum_{n=1}^{\infty}  (\mp 1)^{n}
{ K_{1}(an)\,  \vrule \,}_{a = \frac{m}{T}}.
\label{N-N=}
\end{equation}

At the same time, the results (\ref{i-1-1=}) and (\ref{i-0a1=}) 
lead to the following expression for 
the ratio ${\eta}_{x}^{\pm}$  
\begin{equation}
{\eta}_{x}^{\pm}
=  \frac{\Delta m}{T} \, 
{  \frac{\sum_{n=1}^{\infty} (\mp 1)^{n} K_{1}
(a  n)}{ 
\sum_{n=1}^{\infty} \frac{(\mp 1)^{n}}{n} K_{2}
(a n) } \, \vrule \,}_{a = \frac{m}{T}}.
\label{Lambda=}
\end{equation}
(The functions $K_{1}(x), K_{2}(x)$, which appear in the last two 
formuale, are the modified Bessel functions \cite{Gradstein}). 

Relations (\ref{N-N=}) and (\ref{Lambda=}) are rather 
inconvinient. So discussing the problem whether the effect described
at the end of Sec. 3 by formula (\ref{m11-m22}) 
is able to  generate a significant
contribution to the 
observed baryon--antibaryon asymmetry or not, 
it is sufficient to consider the simplest lower and upper bounds for 
the ratio  ${\eta}_{x}^{\pm}$. 

The estimations given in Appendix B lead to the 
conclusion that the ratio ${\eta}_{x}^{\pm}$ 
(\ref{Lambda}) has the follwing simplest (and rather rough) lower and 
upper bounds 
\begin{equation}
{\eta}_{x, min}^{\pm} \, < \,
{\eta}_{x}^{\pm} \; < \; {\eta}_{x, mx}^{\pm},
\label{<Lambda<}
\end{equation}
where
\begin{equation}
{\eta}_{x, mx}^{\pm} \,
\stackrel{\rm def}{=} \,
\frac{\Delta m}{T} \, 
{\frac{[ - I_{\Delta N,mx}^{\pm}(a)]}{I_{N,min}^{\pm}(a)} \,
\vrule \,}_{a = \frac{m}{T}},
\label{Lambda-mx}
\end{equation}
and
\begin{equation}
{\eta}_{x, min}^{\pm} \,
\stackrel{\rm def}{=} \,
\frac{\Delta m}{T} \, 
{\frac{[ - I_{\Delta N,min}^{\pm}(a)]}{I_{N,mx}^{\pm}(a)} \,
\vrule \,}_{a = \frac{m}{T}},
\label{Lambda-min}
\end{equation}

So, in the case of fermions (baryons) 
the bounds (\ref{-i-1-1mx}) and (\ref{i-0-min}) yield
\begin{equation}
{\eta}_{x, mx}^{-} \, = \,
\frac{\Delta m}{T} \frac{1}{ (1 - e^{-a})^{2} } 
{\frac{ K_{1}(a) }{ K_{2}(a)} \, \vrule \,}_{a = \frac{m}{T} },
\label{Lambda-mx=}   
\end{equation}
which, within the use of (\ref{K-z<1}), (\ref{K-z>1})
gives for $\frac{m}{T} \ll 1$ 
\begin{equation}
{\eta}_{x, mx}^{-} 
\simeq \frac{\Delta m}{2 m},
\label{Lambda-mx<1}
\end{equation}
and for $\frac{m}{T} \gg 1$,
\begin{equation}
{\eta}_{x, mx}^{-}
\simeq \frac{\Delta m}{T}. 
\label{Lambda-mx>1}
\end{equation}
Analogously, taking into account  (\ref{-i-1-1min}) and (\ref{i-0-mx}) 
one finds
\begin{equation}
{\eta}_{x, min}^{-} \, = \,
\frac{\Delta m}{T}  (1 - e^{-a})^{2} 
{\frac{ K_{1}(a) }{ K_{2}(a)} \, \vrule \,}_{a = \frac{m}{T} }.
\label{Lambda-min=}   
\end{equation}
From this expression and from (\ref{K-z<1}), (\ref{K-z>1})
it follows 
\begin{eqnarray}
{\eta}_{x, min}^{-} & \simeq & \frac{1}{2} \frac{\Delta m}{T}
\Big( \frac{m}{T} {\Big)}^{3}, \; \; \; \; (
{\scriptstyle \frac{m}{T} \ll 1}),
\label{Lambda-min<1} \\
{\eta}_{x, min}^{-} & \simeq & \frac{\Delta m}{T} , \; \; \; \;
({\scriptstyle \frac{m}{T} \gg 1}).
\label{Lambda-min>1}
\end{eqnarray}

Considering the case of bosons and using (\ref{+i-1-1mx}) and
(\ref{i-0+min}) one obtains 
\begin{equation}
{\eta}_{x, mx}^{+} \, = \,
\frac{\Delta m}{T} (1 + e^{-a})^{2} 
{\frac{ K_{1}(a) }{ K_{2}(a)} \, \vrule \,}_{a = \frac{m}{T} }.
\label{Lambda+mx=}   
\end{equation}
Thus, keeping in mind (\ref{K-z<1}), (\ref{K-z>1}), the following 
conclusions can be drawn
\begin{eqnarray}
{\eta}_{x, mx}^{+} & \simeq & 2 \frac{\Delta m}{T}
\frac{m}{T},  \; \; \; \; (
{\scriptstyle \frac{m}{T} \ll 1}),
\label{Lambda+mx<1} \\
{\eta}_{x, mx}^{+} & \simeq & \frac{\Delta m}{T} , \; \; \; \;
({\scriptstyle \frac{m}{T} \gg 1}).
\label{Lambda+mx>1}
\end{eqnarray}
The lower bound for ${\eta}_{x}^{+}$ can be found using 
(\ref{+i-1-1min}) and (\ref{i-0+mx}). One has
\begin{equation}
{\eta}_{x, min}^{+} \, = \,
\frac{\Delta m}{T}  \frac{1}{ (1 + e^{-a})^{2} }
{\frac{ K_{1}(a) }{ K_{2}(a)} \, \vrule \,}_{a = \frac{m}{T} }.
\label{Lambda+min=}   
\end{equation}
By means of
(\ref{K-z<1}), (\ref{K-z>1}) one can obtain
the following estimations for 
${\eta}_{x, min}^{+}$ 
\begin{eqnarray}
{\eta}_{x, min}^{+} & \simeq & \frac{1}{8} \frac{\Delta m}{T}
\frac{m}{T},  \; \; \; \; (
{\scriptstyle \frac{m}{T} \ll 1}),
\label{Lambda+min<1} \\
{\eta}_{x, min}^{+} & \simeq & \frac{\Delta m}{T} , \; \; \; \;
({\scriptstyle \frac{m}{T} \gg 1}).
\label{Lambda+min>1}
\end{eqnarray}

Now let us consider, e.g.,  the case $\frac{m}{T} \gg 1$. 
One observes that in this case
\begin{equation}
{\eta}_{x, mx}^{\pm} = {\eta}_{x, min}^{\pm} \equiv {\eta}_{x}
\simeq \frac{\Delta m}{T}. \label{L>1}
\end{equation}
One can convert  this formula into the following one
\begin{equation}
\frac{\Delta m}{m} \, = \, {\eta}_{x} \, \frac{T}{m}, \; \; 
(m \gg T), \label{d-m-m}
\end{equation}
which is valid for unstable fermions (baryons or leptons) as 
well for bosons (mesons).

Theory of big bang nucleosynthesis, which very accurately predicts
the abundances of all light elements, as well as
the comparision of baryon and foton densities in the Universe,
lead to the conlusion that the there should be the following bounds
for the ratio ${\eta}_{x}$: $ 10^{-9} \geq {\eta}_{x} \geq 10^{-10}$ 
\cite{m-antim} --- \cite{m-antim1}. So, using the relations 
(\ref{L>1}), or (\ref{d-m-m}), 
and inserting there, eg.,  ${\eta}_{x} = {\eta}_{x, mx} =
10^{-9}$ one can obtain some allowed, numerical value for 
$\Delta m$. Thus one can verify if 
the effect described by the relation (\ref{m11-m22}) can produce
suitable $\Delta m$. From the formula (\ref{m11-m22}) it follows 
that if $H_{0}$ and $|H_{12}|$ are suitable large and 
$|H_{12}| \sim |H_{0}|$, or even $|H_{12}| > |H_{0}|$ then 
the expression (\ref{m11-m22}) for $\Delta m \equiv 
\Re \, (h_{11}^{\mit \Theta} - h_{22}^{\mit \Theta})$ 
can take the required values. Indeed, expanding 
${\Sigma}_{12}(H_{0} \pm |H_{12}|)$ by its Taylor series expansion
one finds (to the lowest order of $|H_{12}|$)
\begin{equation}
\Delta m \equiv \Re \, (h_{11}^{\mit \Theta} - h_{22}^{\mit \Theta})
= 2\, \Im \, \Big( H_{21}
{ \frac{\partial {\Sigma}_{12}^{I}(x)}{\partial x} 
\, \vrule \,}_{x=H_{0}}
\, \Big) \, + \, \ldots  \,.
\label{d-m-T}
\end{equation}
One should stress that due to the presence of resonace terms, the 
derivative $\frac{\partial {\Sigma}_{12}^{I}(x)}{\partial x}$
need not be small, and so the product 
$H_{21} \frac{\partial {\Sigma}_{12}^{I}(x)}{\partial x}$ in
(\ref{d-m-T}), especially if $|H_{12}|$ is suitable large. Note 
that the contribution of the higher terms of Taylor series 
expansion of ${\Sigma}_{12}(H_{0} \pm |H_{12}|)$ into $\Delta m$ 
can be even more large than (\ref{d-m-T}) for suitably large  
$|H_{12}|$.  
The conditions guaranteeing the required magnitude of
$|H_{12}|$ can be met at the first instants of the exsistence of our 
Universe. (The estimations (\ref{FL}), (\ref{hz-h0}) refer to the 
today's 
epoch and to the case of neutral K mesons only).  Note also that 
in the relation (\ref{d-m-m})
the parameter ${\eta}_{x}$  is multiplied by very smal factor 
$\frac{T}{m}$. 
Therefore it seems that 
the effect described by formulae (\ref{m11-m22}) and (\ref{d-m-T}) 
can have a significant contribution to 
the observed baryon --- antibaryon asymmetry, and thus it is able, 
maybe together 
with one from the other mechanisms considered in \cite{m-antim}
--- \cite{m-antim13},  to explain the matter ---
antimatter asymmetry. A more detailed analysis of the 
cosmological consequences of the property (\ref{m11-m22}) can be
given in the future papers.
 
Note that the magnitude of $|H_{12}|$ and $|H_{0}|$ has not any 
effect on the validity of the approximate formulae (\ref{B10})
for matrix elements $v_{jk}$ of $V_{\parallel} 
\simeq V_{\parallel}^{(1)}$, by means of which matrix elements
$h_{jk}$ of $H_{\parallel}$ are defined. The condition (\ref{B6}),
3which secures  the validity of this approximation, does not depend
on $|H_{12}|$. So, the effect expressed by the relation
(\ref{m11-m22}) occurs for large $|H_{12}|$ too.  

\section{Final remarks.}

The approximation described in Sec. 2.2 shows that at $t = 0$ the 
diagonal matrix  elements $h_{11}(0), h_{22}(0)$ of the effective 
Hamiltonian $H_{\parallel}(t)$ are equal, which means that masses
of particles nad their antiparticles are equal at the instant of 
their creation. The property (\ref{m11-m22}) occurs for 
$t > T_{as}^{X} > 0$, (here $T_{as}^{X}$ is calculated 
for the pairs of particles $X, \overline{X}$ discussed in the 
previous Section and it corresponnds to the time 
$T_{as}$, which was calculated for the neutral K system),
which means (as it follows from Sec. 5.2) that if 
the baryon number B is not conserved and if CP symmetry is violated 
then the asymmetry between the numbers 
of unstable baryons and antibaryons can arise in a CPT invariant system
at $t > T_{as}^{X} > 0$ even in the thermal equilibrium state of this 
system. This means that the effect described in Sec. 3 can 
replace the major part of
models of type considered in \cite{m-antim11} --- \cite{m-antim1a} 
and it has major advantage over those models, namely  
it does not require CPT symmetry to be violated.

In this place the one more problem should be touched. Namely
the quantum field theory provides us with the recipe of how
to construct some charge and number operators. The baryon
number operator $\hat{B}$ is used in some papers 
in order to prove that the third Sakharov condition for bryogenesis
is necessary.  Namely in these papers assuming (\ref{l19}) the 
equilibrium average of baryons, $<B>_{T}$, was calulated 
\begin{eqnarray}
<B>_{T} & = & {\rm Tr} \, (e^{-\beta H} \hat{B}) \; = \;
{\rm Tr} \, \Big({\Theta}^{-1} {\Theta} (e^{- \beta H} \hat{B} ) \Big) 
\nonumber \\
& \equiv & 
{\rm Tr} \, \Big( {\Theta} (e^{- \beta H} \hat{B}) {\Theta}^{-1} \Big)
\; = - <B>_{T}.  \label{TrB}
\end{eqnarray}
From this equation it follows that $<B>_{T} = 0$, which is considered 
in the large literature as the proof that there cannot be any 
generation of the net of baryon number in equilibrium.  Such a 
conclusion
is true, but only for stable baryons (generally, only for stable
particles). This is because such operators as the baryon number operator 
(and in general any charge operator) are expressed by means of fields
corresponding to the particles considered. There are no quantum 
fields which correspond to unstable particles. Therefore 
any charge operators for unstable particles do not exist. 
Simply, calculations leading to the relation (\ref{TrB}) cannot be
performed in the case of unstable particles. This means that the 
conclusions following from the relation (\ref{TrB}) can not be extended
to the case of unstable elementary quantum objects which are considered 
in this paper. Mathematics and logic give no reasons for which 
consequences of the relation (\ref{TrB}) can be extrapolated to the case
of unstable particles.

It seems that estimations (\ref{FL}), (\ref{hz-h0}) should be 
also taken into 
account when one wants to interepret tests of quantum mechanics and CPT 
symmetry in the neutral kaon system \cite{qm-viol,qm-viol1}. 
Indeed, the parameters 
used in \cite{qm-viol} to describe the deviations of quantum 
mechanics, or violations of CPT, are of similar order to  
(\ref{FL}), (\ref{hz-h0}) \cite{qm-viol1,hayakawa} . This means that the 
interpretation of CPT tests, or tests of modified quantum mechanics, 
based on the theory developed in \cite{qm-viol,qm-viol1} may be 
incorretct. A similar conclusion seems to be 
right with reference to the theories describing effects of external 
fields on the neutral kaon system \cite{5th-f}.  Also, the 
interpretation of tests of special relativity and of the equivalence 
principle \cite{sr-eqp} is based on the standard form, (\ref{b5}), 
(\ref{b5a}) of the $H_{LOY}$. The order of the effects discussed in 
\cite{sr-eqp} can be compared to (\ref{FL}). So it seems to be obvious 
that  the application of $H_{\parallel}$, (\ref{l7b}), (\ref{B13}), 
(\ref{B15}) instead of $H_{LOY}$, when one considers theories developed 
in all these papers, can lead to conclusions which need not agree with 
those obtained in \cite{qm-viol}--- \cite{sr-eqp}. 

Finishing our considerations one ought to mention one more property
of relations (\ref{h11-h22}) and (\ref{m11-m22}). Namely, if instead 
of (\ref{cp}), one has $[{\cal C}{\cal P}, H] = 0$, then 
$h_{11} - h_{22} = 0$ in (\ref{h11-h22}) and $M_{11} - M_{22} = 0$ in 
(\ref{m11-m22}) for stable and unstable states $|{\bf 1}>, |{\bf 2}>$ 
\cite{is}.

Last of all it should be emphasized that the relations obtained in 
Sec. 3 and conclusions (\ref{h11-h22}), (\ref{m11-m22}) and others 
following from them are not a hypothesis. This is a rigorous 
mathematical result obtained within the use of basic assumptions of 
Quantum Mechanics.

\hfill \\
\renewcommand{\theequation}%
{\Alph{section}.\arabic{equation}}
\appendix
\section{Appendix}
\setcounter{equation}{0}

Performing the integration by parts in (\ref{i-0a1}) 
one finds 
\begin{eqnarray}
I_{N}^{\pm}(a) & \equiv &
\frac{1}{3}
\int_{a}^{\infty} (x^{2} - a^{2} )^{\frac{3}{2}}
\frac{e^{x}}{(e^{x} \pm 1)^{2} } dx
\label{i-0a} \\
& = & \frac{1}{3}
\int_{a}^{\infty} (x^{2} - a^{2} )^{\frac{3}{2}}
\frac{e^{-x}}{(1 \pm e^{-x} )^{2} } dx,
\label{i-0aa} 
\end{eqnarray}
where $a>0$.

The function $I_{\Delta N}^{\pm}(z)$, (\ref{i-1-1}), is defined as 
follows
\begin{eqnarray}
I_{\Delta N}^{\pm}(z) & \stackrel{\rm def}{=} &
{ \frac{\partial I_{N}^{\pm}(x)}{\partial x} \, \vrule \,}_{x=z} 
\nonumber \\& = &
- z \int_{z > 0}^{\infty}
(x^{2} - z^{2})^{\frac{1}{2}}
\frac{e^{x}}{(e^{x} \pm 1)^{2}} dx 
\label{i-1-1a} \\
& = &
- z \int_{z > 0}^{\infty}
(x^{2} - z^{2})^{\frac{1}{2}}
\frac{e^{-x}}{(1 \pm e^{-x})^{2}} dx .
\label{i-1-1aa} 
\end{eqnarray}

Now let us consider the integral 
\begin{equation}
{\cal J}^{\pm}_{l} (z)  \stackrel{\rm def}{=} 
\int_{z > 0}^{\infty}
(x^{2} - z^{2})^{\frac{l}{2}}
\frac{e^{-x}}{(1 \pm e^{-x})^{2}} dx .
\label{j-l}
\end{equation}
This integral can be transformed into the following one
\begin{eqnarray}
{\cal J}^{\pm}_{l} (z) &=&
z^{l + 1} \, 
\int_{1}^{\infty}
(x^{2} - 1)^{\frac{l}{2}}
\frac{e^{-zx}}{(1 \pm e^{-zx})^{2}} dx ,
\label{j-l1} \\
& \equiv & 
z^{l + 1} \, \sum_{n = 1}^{\infty} ( \mp 1)^{n} n \, 
\int_{1}^{\infty}
(x^{2} - 1)^{\frac{l}{2}}
e^{-nzx} \, dx ,
\label{j-l2}
\end{eqnarray}
(where $ z > 0$) which can be intergrated term by term.
This leads to the following result
\begin{equation}
{\cal J}^{\pm}_{l} (z) \, = \, \mp
\frac{ 2^{\frac{l + 1}{2}} }{ \sqrt{\pi}} \, 
{\Gamma}({\scriptstyle  \frac{l + 2}{2}} )
\, a^{\frac{l+1}{2}} \,
\sum_{n=1}^{\infty}( \mp 1)^{n} \, 
n^{- \frac{l-1}{2}} \, 
K_{\frac{l+1}{2}} (nz),
\label{j-ll}
\end{equation}
where ${\Gamma}(x)$ is the Gamma function and $K_{n}(x)$ is the modified
Bessel function (see, eg, \cite{Gradstein}), 
\begin{equation}
K_{\frac{l+1}{2}} (z) =
\Big( \frac{z}{2} {\Big)}^{\frac{l+1}{2}}
\frac{ \sqrt{\pi}}{\Gamma ( {\scriptstyle \frac{l}{2} + 1} )} 
\int_{1}^{\infty} \, 
e^{-zt} (t^{2} - 1)^{\frac{l}{2}} \, dt,  \; \; ( z > 0).
\label{K}
\end{equation}
Thus the integrals 
$I^{\pm}_{N}(a)$, (\ref{i-0a1}), (\ref{i-0aa}) 
and $I^{\pm}_{\Delta N}(a)$, 
(\ref{i-1-1}), (\ref{i-1-1aa}), equal 
\begin{eqnarray}
I^{\pm}_{N}(a) & = & \frac{1}{3} {\cal J}^{\pm}_{3}(a) \nonumber \\
& = &  \mp a^{2} \sum_{n = 1}^{\infty} \frac{(\mp 1)^{n}}{n}
K_{2}(an) , \label{i-0a1=} \\
I^{\pm}_{\Delta N}(a) & = & -a {\cal J}^{\pm}_{1}(a) \nonumber \\
& = & \pm a^{2} \sum_{n = 1}^{\infty} (\mp 1)^{n}
K_{1}(an) . \label{i-1-1=}
\end{eqnarray}

The series representations as well as the asymptotic expansions  
of $K_{1}(z)$, $K_{2}(z)$ for $z \ll 1$ and for $z \gg 1$ can be found, 
e.g.,in \cite{Gradstein}. Unfortunately the mentioned asymptotic 
expansions are useless when one tries to estimate the infinite series 
appearing in (\ref{i-0a1=}), (\ref{i-1-1=}). This can be achieved if to 
use the following more simple estimations. Namely,
using relation (\ref{K}) one can majorize the
modified Bessel function, eg., as follows
\begin{equation}
K_{\frac{l+1}{2}} (z) \, < \,
K_{\frac{l+1}{2}}^{up} (z) 
\stackrel{\rm def}{=} 
\Big( \frac{z}{2} {\Big)}^{\frac{l+1}{2}}
\frac{ \sqrt{\pi}}{\Gamma ( {\scriptstyle \frac{l}{2} + 1} )} 
\int_{1}^{\infty} \, 
e^{-zt} t^{l} \, dt,  \; \; ( z > 0).
\label{K-up0}
\end{equation}
The function  $K_{\frac{l+1}{2}}^{up} (z)$ equals
\begin{equation}
K_{\frac{l+1}{2}}^{up} (z > 0) = 
\Big( \frac{z}{2} {\Big)}^{\frac{l+1}{2}}
\frac{ \sqrt{\pi}}{\Gamma ( {\scriptstyle \frac{l}{2} + 1} )} 
e^{-z} \Big\{ \frac{1}{z} + 
\sum_{k=1}^{l}
\frac{l(l-1) \cdot \ldots \cdot (l - k + 1)}{z^{k+1}}
\Big\} .
\label{K-upl}
\end{equation}

The simplest lower bound $K_{\frac{l+1}{2}}^{low} (z)$ 
for $K_{\frac{l+1}{2}} (z > 0)$ can be found 
analogously. From (\ref{K}) one finds that 
\[
K_{\frac{l+1}{2}} (z > 0) \, > \,
K_{\frac{l+1}{2}}^{low} (z > 0),
\]
where
\begin{eqnarray}
K_{\frac{l+1}{2}}^{low} (z > 0) 
&\stackrel{\rm def}{=} &
\Big( \frac{z}{2} {\Big)}^{\frac{l+1}{2}}
\frac{ \sqrt{\pi}}{\Gamma ( {\scriptstyle \frac{l}{2} + 1} )} 
\int_{1}^{\infty} 
e^{-zt}(t - 1)^{l} \, dt
\label{K-low0} \\
& = &
\frac{\sqrt{\pi}}{ 2^{\frac{l+1}{2}} }
\frac{l!}{\Gamma (\frac{l}{2} + 1) }
\frac{e^{-z}}{z^{\frac{l+1}{2}}} . \; \; (z > 0)
\label{K-low1}  
\end{eqnarray}

From (\ref{K-upl}) one infers 
that functions $K_{1}^{up}(z)$ and $K_{2}^{up}(z)$, 
which majorize $K_{1}(z)$ and $K_{2}(z)$ 
appearing in formuale (\ref{i-1-1=}) and (\ref{i-0a1=}),
are equal to
\begin{eqnarray}
K_{1}^{up}(z >  0) &=& e^{-z} ( 1 + \frac{1}{z}),
\label{K1-up} \\
K_{2}^{up}(z >  0) &=& 
\frac{e^{-z}}{3} ( z + 3 + \frac{6}{z} + \frac{6}{z^{2}}).
\label{K2-up}
\end{eqnarray}
These relations enable us to find 
\begin{eqnarray}
{K_{1}^{up}(z > 0)\, \vrule \,}_{z \ll 1} & \simeq &
\frac{1}{z} , \label{K1-up<1} \\
{K_{1}^{up}(z )\, \vrule \,}_{z \gg 1} & \simeq &
e^{-z} , \label{K1-up>1}
\end{eqnarray}
and
\begin{eqnarray}
{K_{2}^{up}(z > 0)\, \vrule \,}_{z \ll 1} & \simeq &
\frac{2}{z^{2}} , \label{K2-up<1} \\
{K_{2}^{up}(z )\, \vrule \,}_{z \gg 1} & \simeq &
\frac{1}{3} e^{-z} z. 
\label{K2-up>1}
\end{eqnarray}
From (\ref{K-low1}) one obtains
\begin{eqnarray}
K_{1}^{low}(z >  0) &=& \frac{e^{-z}}{z},
\label{K1-low} \\
K_{2}^{low}(z >  0) &=& 
2 \frac{e^{-z}}{z^{2}},
\label{K2-low}
\end{eqnarray}
which means that 
\begin{eqnarray}
{K_{1}^{low}(z >  0)\, \vrule \,}_{z \ll 1} &=& \frac{1}{z}, 
\label{K1-low<1} \\
{K_{2}^{low}(z >  0) \, }_{z \ll 1} &=& 
\frac{2}{z^{2}}.
\label{K2-low<1}
\end{eqnarray}

\section{Appendix}
\setcounter{equation}{0}
The answer for the question if the effect described in Sec. 3 is able to  
generate a such particle--antiparicle masses diference, which  has 
significant contribution into the observed matter--antimatter asymmetry,
can be found using the simplet estimations of the integrals 
$I^{\pm}_{N}({\scriptstyle \frac{M}{T}})$ and 
$I^{\pm}_{\Delta N}({\scriptstyle \frac{M}{T}})$. In order to do this 
the use of the best estimations for the modified Bessel 
functions appearing in (\ref{i-0a1=}), (\ref{i-1-1=}) 
is not necessary. The sufficient estimations can be obtained directly 
from (\ref{i-0aa}) and (\ref{i-1-1aa}).
It is easily to find that 
\begin{equation}
- I^{\pm}_{\Delta N,min}(z > 0) \, < \,
- I^{\pm}_{\Delta N}(z > 0) \, < \, 
- I^{\pm}_{\Delta N,mx}(z > 0), \label{<-i+-<}
\end{equation}
where
\begin{eqnarray}
- I^{-}_{\Delta N,min} (z > 0) & \stackrel{\rm def}{=} &
z \int_{z > 0}^{\infty} \, (x^{2} - z^{2} )^{\frac{1}{2}}\, 
e^{-x}\, dx \, \equiv \, z^{2} K_{1}(z), \label{-i-1-1min} \\
- I^{-}_{\Delta N,mx} (z > 0) & \stackrel{\rm def}{=} &
\frac{z}{(1 - e^{-z})^{2}}
\int_{z > 0}^{\infty} \, (x^{2} - z^{2} )^{\frac{1}{2}}\, 
e^{-x} \, dx \nonumber \\
& \equiv &
\frac{z^{2}}{(1 - e^{-z})^{2}} K_{1}(z) ,
\label{-i-1-1mx}
\end{eqnarray}
and
\begin{eqnarray}
- I^{+}_{\Delta N,min} (z > 0) & \stackrel{\rm def}{=} &
\frac{z}{(1 + e^{-z})^{2}} 
\int_{z > 0}^{\infty} \, (x^{2} - z^{2} )^{\frac{1}{2}}\, 
e^{-x}\, dx  \nonumber \\
& \equiv & \frac{z^{2} }{(1 + e^{-z})^{2}}
K_{1}(z), \label{+i-1-1min} \\
- I^{+}_{\Delta N,mx} (z > 0) & \stackrel{\rm def}{=} &
z \int_{z > 0}^{\infty} \, (x^{2} - z^{2} )^{\frac{1}{2}} \, 
e^{-x} \, dx \,
\equiv \, {z}^{2} K_{1}(z) .
\label{+i-1-1mx}
\end{eqnarray}
Treating  integral (\ref{i-0aa}) analogously yields
\begin{equation}
I^{\pm}_{ N,min}(z > 0) \, < \,
I^{\pm}_{\Delta N}(z > 0) \, < \, 
I^{\pm}_{\Delta N,mx}(z > 0), \label{<-i0+-<}
\end{equation}
where
\begin{eqnarray}
I^{-}_{N,min} (z > 0) & \stackrel{\rm def}{=} &
\frac{1}{3} \int_{z > 0}^{\infty} \, (x^{2} - z^{2} )^{\frac{3}{2}}\, 
e^{-x}\, dx \, \equiv \, z^{2} K_{2}(z), \label{i-0-min} \\
I^{-}_{N,mx} (z > 0) & \stackrel{\rm def}{=} &
\frac{1}{3(1 - e^{-z})^{2}}
\int_{z > 0}^{\infty} \, (x^{2} - z^{2} )^{\frac{3}{2}}\, 
e^{-x} \, dx \nonumber \\
& \equiv &
\frac{z^{2}}{(1 - e^{-z})^{2}} K_{2}(z) ,
\label{i-0-mx}
\end{eqnarray}
and
\begin{eqnarray}
I^{+}_{N,min} (z > 0) & \stackrel{\rm def}{=} &
\frac{1}{3(1 + e^{-z})^{2}} 
\int_{z > 0}^{\infty} \, (x^{2} - z^{2} )^{\frac{3}{2}}\, 
e^{-x}\, dx  \nonumber \\
& \equiv & \frac{z^{2} }{(1 + e^{-z})^{2}}
K_{2}(z), \label{i-0+min} \\
I^{+}_{N,mx} (z > 0) & \stackrel{\rm def}{=} &
\frac{1}{3} \int_{z > 0}^{\infty} \, (x^{2} - z^{2} )^{\frac{3}{2}} \, 
e^{-x} \, dx \,
\equiv \, {z}^{2} K_{2}(z) .
\label{i-0+mx}
\end{eqnarray}
Analyzing the series representations for $K_{1}(z)$ and $K_{2}(z)$ one 
concludes that the leading terms for $ z \ll 1$ \cite{Gradstein} are
\begin{eqnarray}
K_{1}(z) & \simeq & \frac{1}{z}, \; \; (z \ll 1), \nonumber \\
K_{2}(z) & \simeq & \frac{2}{z^{2}}, \; \; (z \ll 1),
\label{K-z<1}
\end{eqnarray}
and that for $z \gg 1$ the leading term of the modified Bessel functions
is \cite{Gradstein}
\begin{equation}
K_{n}(z) \simeq  \sqrt{\frac{\pi}{2z}}
e^{-z}, \; \; (z \gg 1, \; n=1,2, \ldots). \label{K-z>1}
\end{equation}

\section{Appendix}
\setcounter{equation}{0}

For completeness, one more estimation of the integral, 
$I_{N}^{\pm}(z)$, will be considered.
Namely, taking into account that $ x \geq a > 0$  one finds from 
(\ref{i-0a1}) that the simplest upper  bound of $I^{\pm}_{N}(z)$ for $ 
z > 0$ can be chosen as follows
\begin{equation}
I_{N}^{\pm}(z) \; < \; {\Upsilon}^{\pm}(z)
\stackrel{\rm def}{=}
\int_{z }^{\infty} 
\frac{x^{2}}{e^{x} \pm 1} \, dx 
\, \leq \, \int_{0 }^{\infty} 
\frac{x^{2}}{e^{x} \pm 1} \, dx  .
\label{i-N,sup,a}
\end{equation}
This means that (see \cite{Gradstein})
\begin{equation}
{\Upsilon}^{+}(z) \leq \frac{3}{4} {\Gamma}(3) {\zeta}(3) = 
\frac{3}{2} {\zeta}(3), \label{i-N,sup,a+}
\end{equation}
and
\begin{equation}
{\Upsilon}^{-}(z) \leq  {\Gamma}(3) {\zeta}(3) = 
2 {\zeta}(3). \label{i-N,sup,a-}
\end{equation}
Here $\zeta (3)$ is the Riemann's Zeta function of 3.
These estimations are valid for any $z > 0$ but in the literature 
are considered as approximate values of the integral (\ref{i-0a1})
for relativistic particles, $ z = \frac{m}{T} \gg 1$, only  (see,eg., 
\cite{Sarkar}). 

\section{Appendix}
\setcounter{equation}{0}
If one considers, eg., the estimation (\ref{i-N,sup,a-}) as the exact 
value of the integral $I_{N}^{-}(z)$ then one obtains
the following expressions for 
${\eta}_{x, min}^{-}$ and ${\eta}_{x, mx}^{-}$,
\begin{equation}
{\eta}_{x, mx}^{-} = \frac{\Delta m}{T} 
\frac{ [-I_{\Delta N,mx}^{-}(z)] }{2\zeta (3)} =
\frac{1}{2 \zeta (3)} 
{ \frac{z^{2} }{ (1- e^{-z})^{2} } K_{1}(z) \,
\vrule \,}_{z = \frac{m}{T} } , \label{L-mx}
\end{equation}
and
\begin{equation}
{\eta}_{x, min}^{-} = \frac{\Delta m}{T} 
\frac{ [-I_{\Delta N,min}^{-}(z)] }{2\zeta (3)} =
{ \frac{z^{2} K_{1}(z) }{2 \zeta (3)}  
\,\vrule \,}_{ z = \frac{m}{T} } . \label{L-min}
\end{equation}
These relations and, for instance, (\ref{K-z<1}) give
\begin{equation}
{\eta}_{x, mx}^{-} \simeq \frac{1}{2 \zeta (3)} \frac{\Delta m}{m}
\sim \frac{\Delta m}{m} , \; \; \; ({\scriptstyle \frac{m}{T} \ll 1}),
\label{L-mx<1}
\end{equation}
and
\begin{equation}
{\eta}_{x, min}^{-} \simeq \frac{1}{2 \zeta (3)} \frac{\Delta 
m}{T}\, \frac{m}{T} \,
\sim \frac{\Delta m}{T} 
\, \frac{m}{T}, \; \; \; ({\scriptstyle \frac{m}{T} \gg 1}).
\label{L-mx>1}
\end{equation}


\begin{thebibliography}{10} \vspace*{-10pt}
\bibitem{1} 
T. D. Lee, R. Oehme  and  C.  N.  Yang,  Phys.  Rev.,
{\bf 106}, (1957) 340.   \vspace*{-10pt}
\bibitem{2} 
T. D. Lee and C. S.  Wu,  Annual  Review  of  Nuclear
Science, {\bf 16}, (1966) 471.
Ed.:  M.  K.   Gaillard   and   M.   Nikolic,   {\em Weak
Interactions}, (INPN et de Physique des Particules,  Paris,  1977) --- 
Chapt. 5, Appendix A.
S. M. Bilenkij, {\em Particles and nucleus}, vol.  1.  No  1
(Dubna 1970), p. 227 [in Russian]. P.  K.  Kabir,  {\em The  CP-puzzle},
Academic Press, \vspace*{-10pt} New York 1968. 
\bibitem{4} 
J. W. Cronin, Rev. Mod. Phys. {\bf 53}, (1981) 373.
J. W. Cronin, Acta Phys. Polon., {\bf B15}, (1984) 419.
V. V. Barmin, et al., Nucl. Phys. {\bf B247}, (1984) 293.
L. Lavoura, Ann. Phys. (NY), {\bf 207}, (1991) 428.
C. Buchanan, et al., Phys. Rev. {\bf D45}, (1992) 4088.
C. O. Dib, and R. D. Peccei, Phys. Rev.,{\bf D46}, (1992)
2265.
R. D. Peccei, CP and  CPT  Violation:  Status  and  Prospects,
Preprint  UCLA/93/TEP/19,  University  of  California,  June
1993.    \vspace*{-10pt}
\bibitem{5} 
E. D. Comins and P. H. Bucksbaum, {\em Weak interactions of
Leptons and Quarks}, (Cambridge University Press, 1983).
T. P. Cheng and L. F. Li, {\em Gauge Theory of Elementary
Particle Physics}, (Clarendon, Oxford 1984).
\vspace*{-10pt}
\bibitem{6} 
Yu. V. Novozhilov, {\em Introduction to the Theory
of Elementary Particles}, (Nauka, Moskow 1972), (in Russian).
W. M. Gibson and B. R. Pollard, {\em Symmetry Principles
in  Elementary  Particle  Physics},  (Cambridge  University  Press,
1976).      \vspace*{-10pt}
\bibitem{dafne} 
L. Maiani, in {\em The Second Da$\Phi$ne Physics Handbook}, 
vol. 1, Eds. L. Maiani, G. Pancheri and N. Paver, SIS --- Pubblicazioni, 
INFN  --- LNF, Frascati, 1995; pp. 3 --- 26.
\vspace*{-10pt}
\bibitem{3} 
L. A. Khalfin,  The  theory   of
$K_{0}$,${\overline K}_{0}$  ($D_{0}$,${\overline D}_{0}$   and
$T_{0}$,${\overline T}_{0}$ mesons beyond the Weisskopf-Wigner
approximation  and  the
CP--problem, preprint LOMI P--4--80,  Leningrad,  February  1980
\vspace*{-10pt}
\bibitem{leonid} 
L. A. Khalfin, Theory of Unstable Neutrino Mixing and the 
17 keV Neutrino Problem, Preprint HU--TFT-92--21, Helsinki, 
June 1992. \vspace*{-10pt}
\bibitem{baldo} 
M. Baldo--Ceolin, Neutron--antineutron oscillation 
experiments, {\em Proceedings of the "International Conference of 
Unified 
Theories and Their Experimental Tests"} --- Venice --- 16--18 March 
1982; Sensitive Search for Neutron--Antineutron Transitions at the Ill 
Reactor, {\em AIP Conference Proceedings No 125 of the Fifth 
International 
Symposium on "Capture Gamma--Ray Spectroscopy and Related Topics"} --- 
Edited by S. Raman,  Knoxville, September 1984, p. 871. \vspace*{-10pt}
\bibitem{horwitz} 
L. P. Horwitz and J. P. Marchand, Helv. Phys.
Acta {\bf 42} (1969) 801. \vspace*{-10pt}
\bibitem{7} 
W. Kr\'{o}likowski and J. Rzewuski, Bull. Acad. Polon.
Sci. {\bf 4} (1956) 19.
W. Kr\'{o}likowski and J. Rzewuski, Nuovo. Cim.
{\bf B 25} (1975) 739 and refernces therein.
K. Urbanowski, Acta Phys. Polon. {\bf B 14} (1983)
485. \vspace*{-10pt}
\bibitem{pra} 
K. Urbanowski, Phys. Rev. {\bf A 50}, (1994) 2847. 
\vspace*{-10pt}
\bibitem{11} 
K. Urbanowski, Phys. Lett. {\bf B 313},
(1993) 374.   \vspace*{-10pt}
\bibitem{14} 
K. Urbanowski, Is  the  new  interpretation  of  some
standard CPT--violation parameters  necessary?,  Preprint  of  the
Pedagogical University, No WSP--IF 94--39, Zielona G\'{o}ra,
May 1994;  CPT transformation properties of the exact effective 
Hamiltonian for neutral kaon complex, Preprint of the Pedagogical 
University, No WSP--IF 96--44, Zielona G\'{o}ra, March 1996;
CPT transformation properties of the exact effective Hamiltonian 
for neutral kaon and similar complexes, Preprint of the Pedagogical 
University, No WSP--IF 97--50, Zielona G\'{o}ra, July 1997 --- 
{\bf hep--ph/9803376}. \vspace*{-10pt}
\bibitem{ww} 
V. F. Weisskopf and E. T. Wigner, Z. Phys. 
{\bf 63} (1930) 54; \vspace*{-10pt} {\bf 65} (1930) 18. 
\bibitem{9} 
K. Urbanowski, Int. J. Mod.  Phys.  {\bf  A 8},  (1993)
3721. \vspace*{-10pt}
\bibitem{10} 
K. Urbanowski, Int. J. Mod.  Phys.  {\bf A 10}, (1995)
1151.     \vspace*{-10pt}
\bibitem{chiu} 
C. B. Chiu and E. C. G. Sudarshan, Phys. Rev.
{\bf D 42} (1990) 3712; E. C. G. Sudarshan, C. B. Chiu and
G. Bhamathi, Unstable Systems in Generalized Quantum Theory,
Preprint DOE-40757-023 and CPP-93-23, University of Texas,
October 1993. \vspace*{-10pt}
\bibitem{is} 
K. Urbanowski,  Int. J. Mod. Phys. {\bf A 13}, (1998) 965. 
\vspace*{-10pt}
\bibitem{why}
K. Urbanowski, Why the LOY model cannot be used for designing 
CPT--invariance tests in neutral kaons systems? On CPT--noninvariance 
of systems with exponentially decaying particles, Preprint of the 
Pedagogical University No WSP--IF 94--38, Zielona G\'{o}ra, May 
1994.\vspace*{-10pt}
\bibitem{tsai} 
T. Mochizuki, N. Hashimoto, A. Shinobori, S. Y. Tsai, Remarks on 
Theoretical Frameworks Describing The Neutral Kaon System, Preprint 
of The Nihon University, No NUP--A--97--14, Tokyo, June 1997. 
\vspace*{-10pt}
\bibitem{kabir}
P. K. Kabir and A. Pilaftsis, Phys. Rev. {\bf A 53}, (1966) 66. 
\vspace*{-10pt}
\bibitem{m-antim} 
A. D. Sakharov, Pis'ma Zh. Eksp. Teor. Fiz., {\bf 5} (1967) 32 [in 
Russian], [JETP Letters {\bf 5} (1967) 24]; A. A. Grib and 
Yu. V. Kryukov, Yad. Fiz. {\bf 48} (1988) 1842 [in Russian]; M. C.
Bento and O. Bertolami, Phys. Lett {\bf B 323} (1994) 373;  A. 
Yamaguchi and A. Sugamoto, Modern Phys. Lett. {\bf A9} (1994) 2599; 
G. R. Farrar and M. E. Shaposhnikov, Phys. Rev. Lett. {\bf 70} (1993) 
2833 and Phys. Rev. {\bf D50} (1994) 774; M. B. Gavela, P. Hernandez, 
J. Orloff and O. Pene, Nucl. Phys. {\bf B 430} (1994) 345, 382; K. A. 
Olive, Bing Bang Baryogenesis, in {\em Proceedings of the 33rd 
International Winter School on Nuclear and Particle Physics "Matter 
Under Extreme Conditions"} --- Schladming (Austria) 1994, Eds. H. Latal 
and W. Schweiger (Springer, Berlin, 1994),
{\bf hep--ph/9404352};
P. Huet and E. Sather, Phys. 
Rev. {\bf D51} (1995) 379; 
W. Buchm\"{u}ller and M. Pl\"{u}macher, Phys. 
Lett. {\bf B389} (1966) 73; 
V. A. Rubakov and M. E. Shaposhnikov, Uspekhi Fizicheskich Nauk, 
{\bf 166}, (1966), 493 and {\bf hep--ph/9603208};
B. de Carlos and J. R. Espinosa, Nucl. Phys. 
{\bf B 503} (1997) 24; 
E. Kh. Akmedov, V. A. Rubakov and A. Yu. Smirnov, Phys. Rev. Lett.,
{\bf 81}, (1998), 1359. 
\vspace*{-10pt}
\bibitem{Sarkar}S. Sarkar, Reports on Progres in Physics, {\bf 59},
(1996), 1493. 
\vspace*{-10pt}
\bibitem{m-antim0}
M. Trodden, Electroweek Baryogenesis, Preprint
No CWRU--P6--98 --- March 1998,
{\bf hep--ph/9803479}. \vspace*{-10pt}
\bibitem{m-antim11}A. D. Dolgov, Ya. B. Zeldovich, Reviews of Modern 
Physics, {\bf 53}, (1981),1. \vspace*{-12pt}
\bibitem{m-antim12}A. G. Cohen, D. B. Kaplan, Physics Letters 
{\bf B 199}, (1987), 251. \vspace*{-12pt}
\bibitem{m-antim13}A. D. Dolgov, Baryogenesis 30 years after, Preprint
No TAC--1997--024 --- July 1997, {\bf hep--ph/9707419}. \vspace*{-12pt}
\bibitem{m-antim1} 
S. Barshay, Z. Phys. {\bf C74} (1997) 709. \vspace*{-10pt}
\bibitem{m-antim1a}
O. Bertolami, Don Colladay, V. Alan Kostelecky and R. Potting, 
Phys. Lett. {\bf B 395} (1997) 178. \vspace*{-10pt}
\bibitem{cpt} 
W. Pauli, in: {\em Niels Bohr and the Developmnet of Physics}, ed. W. 
Pauli (pergamon Press, London, 1955), pp. 30 ---51. G. Luders, Ann. 
Phys. 
(NY) {\bf 2} (1957) 1. . R. Jost, Helv.Phys. Acta {\bf 30} (1957) 409.
R.F. Streater and A. S. Wightman, {\em CPT, Spin, Statistics and All 
That}, (Benjamin, New York, 1964).N. N. Bogolubov, A. A. Logunov and 
I. T. Todorov, {\em Introduction to Axiomatic Field Theory},
\vspace*{-10pt}(Benjamin, New York, 1975).  
\bibitem{messiah} 
A. Messiah, {\em Quantum Mechanics}, vol. 2, (Wiley,New York 1966). 
\vspace*{-10pt}
\bibitem{bohm} 
A. Bohm, {\em Quantum Mechanics: Foundations and Applications}, 2nd ed., 
(Springer, New York 1986). \vspace*{-10pt}
\bibitem{15} 
E. P. Wigner, in: {\em Group Theoretical Conceptsand Methods in 
Elementary 
Particle Physics}, ed.: F. G\"{o}resy,(New York 1964). \vspace*{-10pt}
\bibitem{improved} 
K. Urbanowski and J. Piskorski, Improved Lee, Oehme and Yang 
approximation, Preprint of the Pedagogical University No WSP--IF 
98--51, Zielona G\'{o}ra, March 1998 --- 
{\bf physics/9803030}.\vspace*{-10pt}
\bibitem{data} 
Review of Particle Properties, Physical Review {\bf D 54},(1966) No 1.
\vspace*{-10pt}
\bibitem{urb-pla} 
K. Urbanowski, Phys. Lett. {\bf A 171}, (1992) 151.
\vspace*{-10pt}
\bibitem{qm-viol}
J. Ellis, J. S. Hagelin, D. V. Nanopoulos and M. Srednicki, Nucl. Phys.
{\bf B 241}, (1984), 381;
J. Ellis,  N.E. Mavratos and D.V. Nanopulous, Phys. Lett., {\bf B 293},
(1992), 142;
V.A. Kostelecky, Phys. rev. Lett., {\bf 80}, (1998), 1818.
\vspace*{-10pt}
\bibitem{qm-viol1}
J. Ellis, J. L. Lopez, N. E. Mavromatos and D. V. Nanopoulos ,
Phys. Rev. {\bf D 53}, (1996), 3846. \vspace*{-10pt}
\bibitem{hayakawa}
M. Hayakawa, Measurement of CPT violation in the neutral kaon system,
{\em Proceeding for first KEK meeting on "CP violation and its origin"
in 1993}, {\bf hep--ph/9704418}. \vspace*{-10pt}
\bibitem{5th-f}
E. Fischbach, C. Talmadge, Nature, {\bf 356}, (1992), 207; 
E. Fischbach, D. Sudarsky, A. Szafer, C. Talmadge and
S. H. Aronson,Phys. Rev. Lett., {\bf 56}, (1986), 3; 
D. Sudarsky, E. Fischbach, C. Talmadge, A. H. Aronson and H.--Y.
Cheng, Annals of Physics, {\bf 207}, (1991), 103; 
E. Fischbach, C. Talmadge, Preprint No {\bf hep--ph/9606249}.
\vspace*{-10pt}
\bibitem{sr-eqp}
T. Haymbye, R. B. Mann, U. Sarkar, Phys. Lett., {\bf B421}, (1998),
105; 
T. Haymbye, R. B. Mann, U. Sarkar, Phys. Rev. {\bf D58}, (1998),
art. no 025003; 
C. Alvarez, R. B. Mann, Phys. Rev., {\bf D55}, (1997), 1732;
R. J. Hughes, Phys. Rev., {\bf 46}, (1992), R2283;
I. R. Kenyon, Phys. Lett., {\bf B 237}, (1990), 274. 
\vspace*{-10pt}
\bibitem{Gradstein}I. Gradstein and I. M. Ryshik, {\em Tables of series, 
products and integrals}, (Harri Deutsch, Thun, 1981).
\end{thebibliography}
\end{document}